\documentclass[nofootinbib]{revtex4-2} 
\usepackage[utf8]{inputenc}
\usepackage{longtable}
\usepackage{xcolor}
\usepackage{blindtext,alltt}
\usepackage{amsmath}
\usepackage{mathrsfs}
\usepackage{amsfonts}
\usepackage{amssymb}
\usepackage{soul}
\usepackage{graphicx}
\usepackage{tikz}
\usepackage{bm}
\usetikzlibrary{cd, positioning}

\usepackage{ulem} 

\newcommand{\half}{\frac{1}{2}}
\newcommand{\I}{|}
\renewcommand{\H}{\mathcal{H}}

\newcommand{\e}{\textrm{e}}
\newcommand*\widebar[1]{%
\hspace{0.2em}
  \hbox{%
    \vbox{%
      \hrule  
      \kern0.2ex
      \hbox{%
        \kern-0.15em
        \ensuremath{#1}%
        \kern-0.15em
      }%
    }%
  }%
  \hspace{0.2em}
} 
\newcommand{\erf}{\textrm{erf}}
\newcommand{\erfc}{\textrm{erfc}}
\newcommand{\abs}[1]{\left\I #1 \right\I}
\newcommand{\p}{\mathbf{p}}

\def\Xint#1{\mathchoice
   {\XXint\displaystyle\textstyle{\boldsymbol{#1}}}%
   {\XXint\textstyle\scriptstyle{\boldsymbol{#1}}}%
   {\XXint\scriptstyle\scriptscriptstyle{\boldsymbol{#1}}}%
   {\XXint\scriptscriptstyle\scriptscriptstyle{\boldsymbol{#1}}}%
   \!\int}
\def\XXint#1#2#3{{\setbox0=\hbox{$#1{#2#3}{\int}$}
     \vcenter{\hbox{$#2#3$}}\kern-.5\wd0}}

\def\dashint{\Xint -}

\newcommand{\IFI}[3]{ {}_{1}F_{1}\Big(#1, #2 ; #3 \Big)}
\newcommand{\nCr}[2]{\begin{pmatrix}#1\\#2\end{pmatrix}}
\newcommand{\ierf}[1]{\mathfrak{I}^{#1}\textrm{erfc} }

\newcommand{\R}[1]{#1^{\star}}
\definecolor{grey}{rgb}{0.75, 0.75, 0.75}

\begin{document}
\title{Non-perturbative quantum propagators in bounded spaces}

\author{James P. Edwards}
\email{james.p.edwards@umich.mx}
\affiliation{Instituto de F\'isica y Matem\'aticas,\\
Universidad Michoacana de San Nicol\'as de Hidalgo, Morelia, M\'exico.}

\author{V\'ictor A. Gonz\'alez-Dom\'inguez}
\email{vgonzalez@im.upchiapas.edu.mx}
\affiliation{Universidad Polit\'ecnica de Chiapas, Carretera Tuxtla-Villaflores, Km. 1+500, Las Brisas, 29150 Suchiapa, Chiapas, Mexico.}

\author{Idrish Huet}
\email{idrish.huet@gmail.com}
\affiliation{Facultad de Ciencias en F\'isica y Matem\'aticas,\\Universidad Aut\'onoma de Chiapas,\\ Ciudad Universitaria, Tuxtla Guti\'errez 29050, M\'exico.}

\author{Maria Anabel Trejo}
\email{m.trejo-espinosa@hzdr.de}
\affiliation{
Helmholtz-Zentrum Dresden-Rossendorf,\\Bautzner Landstra\ss e 400
01328, Dresden, Germany.}

\begin{abstract}
We outline a new approach to calculating the quantum mechanical propagator in the presence of geometrically non-trivial Dirichlet boundary conditions based upon a generalisation of an integral transform of the propagator studied in previous work (the so-called ``hit function''), and a convergent sequence of Padé approximants. In this paper the generalised hit function is defined as a many-point propagator and we describe its relation to the sum over trajectories in the Feynman path integral. We then show how it can be used to calculate the Feynman propagator. We calculate analytically all such hit functions in $D=1$ and $D=3$ dimensions, giving recursion relations between them in the same or different dimensions and apply the results to the simple cases of propagation in the presence of perfectly conducting planar and spherical plates. We use these results to conjecture a general analytical formula for the propagator when Dirichlet boundary conditions are present in a given geometry, also explaining how it can be extended for application for more general, non-localised potentials. Our work has resonance with previous results obtained by Grosche in the study of path integrals in the presence of delta potentials \cite{grosche1990path,grosche1993delta,grosche1995delta}.
We indicate the eventual application in a relativistic context to determining Casimir energies using this technique. 
\end{abstract}

\maketitle

\section{Introduction}
The quantum mechanical propagator, $K(y, x; T)$, (also called the kernel) is defined in terms of the configuration space matrix elements of the time evolution operator,
\begin{equation}
    K(y, x; T) = \big \langle y \big \I \e^{-i\widehat{H}T} \big\I x \big \rangle\,  \theta (T)\,, \label{QMpropagator}
\end{equation}
where $\widehat{H} = \frac{\widehat{p}^{2}}{2m} + V(\widehat{x})$ is (stationary) the Hamiltonian of a system with potential $V$. The kernel satisfies the position space Schrödinger equation with the boundary condition $K(y,x; 0^+) =  \delta^D (y-x)$ 
and as such contains the full information about the quantum system. Except in some specific cases, or for systems with special symmetries, finding the explicit form of the kernel can be very non-trivial and even conventional perturbation theory has its limitations, especially for strongly perturbing potentials, so new techniques for its calculation or estimation are of significant interest.

In \cite{PvHz}, two invertible integral transforms of the kernel were introduced. The ``hit function'' defined there is denoted by $\widebar{\mathcal{H}}(z \I y, x ; T)$  and represents the contribution to the kernel involving propagation that passes through (``hit'') the spatial point $z$; it can be used to recover the propagator by integrating over this point:
\begin{align} \label{1hitbar}
    \widebar{\mathcal{H}}(z \I y,  x; T) &= \frac{1}{T}\int_{0}^{T}dt\, K(z, x; t)K(y, z; T-t)\,, \\
    K(y, x; T) &= \int d^{D}z\, \widebar{\mathcal{H}}(z \I y, x; T)\,.
\end{align}
This function also finds application in prior studies of Brownian motion and Levy random walks where it is related to the concept of ``local time'' \cite{PhysRevE.92.062137, PhysRevE.95.052136} that measures the proportion of time for which a particle's trajectory is found in a particular region. For the free particle, which will be the focus of this paper, the  Euclidean (imaginary-time) hit function was calculated in one dimension in \cite{PvHz} to be
\begin{equation}
    \widebar{\mathcal{H}}_{0}(z \I y, x; T) = \frac{m}{2T}\erfc\Big[ \sqrt{\frac{m}{2T}} \big( \abs{x-z} + \abs{z-y} \big) \Big]\,,\quad  T\geq 0\,,
\end{equation}
which was verified by a numerical sampling of the Feynman path integral (note that this function is constant for $z$ between $y$ and $x$. Also $\erfc$ is the complementary error function defined precisely below).

For potentials that are sharply localised, such as\footnote{Here $\delta_{\epsilon}$ indicates a localised, $\delta$-function-like potential that may be smeared over some small characteristic scale $\epsilon$ (it may be extended in a codimension $d > 0$ subspace of $\mathbb{R}^{D}$).} $V(x) \sim \lambda \delta_{\epsilon}(x - a)$, the perturbative expansion of the path integral representation of the kernel (see equation (\ref{QM}) below) in powers of the coupling constant, $\lambda$, naturally constrains the path integral to trajectories that must pass through (or in a neighbourhood of) the point $a$ various times. This is the case, for example, for  many scattering problems and contact interactions \cite{Curry:2017cnu, Edwards:2015hka, Mansfield:2011eq} and varied models of lattice structure in condensed matter \cite{kronig1931quantum, demkov2013zero}. In the relativistic case, constrained path integrals arise in the context of the Casimir effect where one is interested in trajectories that touch the conducting plates \cite{PhysRevD.94.105009,gies1,gies2,gies3}, and for Dirichlet boundary conditions imposed in some spatial region \cite{Clark:1980xt,  Farhi:1989jz, Bastianelli:2006hq, Bastianelli:2007jr, bastianelli2008scalar, Vinas:2010ix, Corradini:2019nbb, Carreau:1991yx, Asorey:2006ij, Asorey:2007zza} where the contributions from paths passing through this region should be removed. A fresh approach to calculating, either analytically, approximately or numerically, such path integrals will therefore find wide application.

The purpose of this article is three-fold. We will introduce the generalisation of the hit function to the ``$n$-hit function'' that counts trajectories that pass through $n$ intermediate spatial points $\{z_{1}, \ldots , z_{n}\}$ \textit{in chronological order} whilst travelling between the endpoints $x$ and $y$. This $n$-hit function can be defined as the constrained path integral (in Euclidean space-time, with $\tau_{n+1} := T$)
\begin{equation}
   \hspace{-0.5em} \mathcal{H}(z_{1}, \ldots z_{n} \I y, x; T) :=\int_{x(0) = x}^{x(T) = y}\hspace{-1.5em}\mathscr{D}x(\tau)\, \prod_{i = 1}^{n}\int_{0}^{\tau_{i+1}} \!d\tau_{i}\,  \delta^{D}\big( x(\tau_{i}) - z_{i}  \big)\, \e^{-\int_{0}^{T}d\tau \big[ \frac{m\dot{x}^{2}}{2} - V(x(\tau))\big]}\, \theta (T) .
   \label{eqHPIdef}
\end{equation}
Unless otherwise indicated we will mean the free case when we refer to the $n$-hit function which is made explicit in expressions via the notation $\mathcal{H}_0$. We will uncover relations between the $n$-hit function for different \textit{order}, $n$, and in different number of spatial dimensions, $D$, and will give explicit formulae that allow it to be determined for arbitrary $n$ and $D$. We will also use this function and the theory of Pad\'e approximants to conjecture a closed analytical formula for the propagator in the presence of Dirichlet boundary conditions in arbitrary geometries and support our claim by applying it to some simple examples.

It is important to highlight some prior studies based on a similar philosophy. The Path Decomposition Expansion of \cite{PhysRevLett.53.411, Halliwell:1995jh, Yearsley_2009, Yearsley_2008} splits up the path integral representation of the Green function associated to the kernel in terms of contributions from trajectories in different spatial regions; in \cite{Koch:2020dql, Koch:2017nha} the Multi-Step Propagator is introduced with the aim of eliminating a possibly dangerous overcounting of certain types of trajectory in the path integral; the effective action of a (relativistic) scalar field in a localised potential is found in \cite{Franchino-Vinas:2020okl} by resumming a perturbative expansion of the interaction (in fact our $n$-hit function is related to an infinite set of intermediate functions that enter this resummation); and finally, Polyakov long ago formulated a representation of the (one-particle-reducible) contributions to the $N$-point configuration-space amplitudes in $\phi^{3}$ theory in terms of a relativistic point-particle path integral constrained to pass through the prescribed external points \cite{Polyakov:1987ez}. 

However, here we work with the path integral representation of the propagator as is, in configuration space, and work out the relative contributions from the (infinite number of) trajectories that pass through the prescribed points. The $n$-hit function we define has a simple physical interpretation according to context: in the quantum mechanical sense it represents the (un-normalised) amplitude of propagation for a particle to travel from point $x$ to point $y$ in time $T$, going through the points $z_1, \cdots, z_n$, (in time order) per unit  $(n+1)$-volume $d^D z_1 \cdots d^D z_n d^D y$ (following the usage in \cite{feynman2010quantum} we will refer to such an amplitude as a {\it relative amplitude}); in the context of Brownian dynamics it describes the probability density of the propagation taking place under the same constraints, per unit $n$-volume of the intermediate points. 

This information can be useful for analytic and numerical estimations of the kernel for systems whose path integral cannot be computed in close form, since it can indicate which trajectories, or regions of space, will be dominant in its determination. In fact the hit function was already sampled in works based on worldline numerics \cite{PvHz} and the extension considered here could shed new light on the undersampling problem encountered in \cite{Edwards:2019fjh, Corradini:2020tgk, Franchino-Vinas:2019udt}. Looking further ahead, the methods we describe here are well-adapted to the problem of estimating the Casimir energy for arbitrary surface geometries, even for partially conducting plates. Indeed, via the lesser-well-known Worldline Formalism of quantum field theory \cite{ChrisRev, UsRep, SchmidtRev},  the generalised hit function defined here admits a clear extension to field theory processes where it will be defined for relativistic point particles in Minkowski space (as an immediate example, the $n$-hit function can be converted into Polyakov's representation of $\phi^{3}$ amplitudes cited above by transforming to Minkowski space and introducing a (Schwinger) integral over the trajectories' proper time \cite{Polyakov:1987ez}). This case shall be addressed in future work --  but see \cite{pisanietal} where the heat kernel with Dirichlet boundary conditions on a $D$-dimensional manifold ($D$-ball) was obtained using worldline techniques and conformal transformations. 

The outline of this paper is as follows: section II introduces the $n$-hit function in terms of path integrals and gives also an $n$-fold integral representation in terms of the free particle kernel $K_0$. Section III presents one of the main results of this work, a novel conjectured representation for the propagator/heat kernel on a bounded region with Dirichlet boundary conditions based on Pad\'e approximants. Section IV.A shows a closed formula for the free $n$-hit function in $D=3$ spatial dimensions, the case of most direct physical relevance and also special for being easily accessible analytically. Section IV.B gives the energy representation of the general $n$-hit function for any space dimension $D$. Section V demonstrates how certain general operations connect $n$-hit functions in different spatial dimensions $D$ and of different orders $n$ to one another. An explicit realisation of these operators is given, providing a set of identities relating the functions that in turn leads to a hierarchy which generates all possible $n$-hit functions. Section VI.A contains numerical results for the propagators inside a sphere in $D=3$ and an infinite plane (also in $D=3$) using the Pad\'e representation proposed in section III.
In section VI.B we obtain an integral energy-representation of the general $n$-hit function in any $D$. Section VII presents our conclusions and insight on further applications of the method proposed and more generally of the $n$-hit function and its connection to Feynman's description of the path integral. Finally, for the sake of clarity most detailed calculations and prior results have been deferred to appendices.

\section{Quantum propagation and the $n$-hit function} \label{QuantP}

The problem we will address is that of quantum propagation in the presence of a potential, and while the methods outlined here can be applied to more general situations we will mainly focus on a special kind of singular potential with support on a predetermined surface, $V(x) = \lambda \int_S \delta (x - z) \,d\sigma $. Here $z$ is a position vector on $S$, $d\sigma$ represents a measure that fully parametrises the surface $S$ and $\lambda$ is a positive\footnote{The repulsive limit $\lambda \to +\infty$ imposes Dirichlet boundary conditions on $S$, while $\lambda<0$ corresponds to an attractive potential on $S$.} constant that fixes the strength of the potential. The propagation amplitude of a non-relativistic particle between position $x$ at time $t=0$ and position $y$ at time $t=T$ in the presence of a potential is known to be given in natural units by the path integral
\begin{equation} \label{QM}
 \langle y ; T | x;0 \rangle  =  \langle y | \e^{-i\widehat{H}T} | x \rangle = \int_{x(0)=x}^{x(T)=y}  \hspace{-1.5em}\mathscr{D} x(\tau)\,
 e^{i \int_0^{T} d\tau \left ( \frac{m}{2}\dot{x}^2 -V(x(\tau))  \right)}\, . 
\end{equation}
Henceforth we will Wick-rotate to the imaginary time formulation of quantum mechanics in which quantum free propagation becomes Brownian motion and all amplitudes become real. Motivated by future applications we will usually henceforth set $m=1/2$ -- if desired the mass $m$ can always be recovered by dimensional analysis in all our results, (the choice $m=1/2$ corresponds to a unit diffusion coefficient) which converts (\ref{QMpropagator}) to

\begin{equation}
   K(y,x;T) = \int_{x(0)=x}^{x(T)=y} \hspace{-1.5em} \mathscr{D} x(\tau)\,
 \e^{- \int_0^T d\tau \left ( \frac{\dot{x}^2}{4} + V(x)  \right)} \,\theta (T).
\end{equation}
A standard way to deal with the localised potential considered here is to expand the exponential to obtain a perturbative series in the constant $\lambda$,
\begin{equation} \label{fullK}
 K(y,x;T) = \int_{x(0)=x}^{x(T)=y} \hspace{-1.5em} \mathscr{D} x(\tau)\,
 e^{- \int_0^T d\tau  \frac{\dot{x}^2}{4} } \sum_{n=0}^{\infty} \frac{(-1)^n}{n!}  \left( \int_0^T V(x(\tau)) d\tau\right)^n \, \theta (T).
\end{equation}
Hence the $n$-th term of this series requires the determination of a constrained path integral
\begin{equation}
   \mathcal{I}_{n}(\{z_{i}\}, \{\tau_{i}\}) := \int_{x(0)=x}^{x(T)=y}\!  \hspace{-1.5em}\mathscr{D} x(\tau)\,
 e^{- \int_0^T d\tau  \frac{\dot{x}^2}{4} }\prod_{j=1}^n \delta^{D} (x(\tau_i) - z_i) \,.
 \label{eqIn}
 \end{equation}
Remarkably, following the steps in \cite{PvHz} one can determine this in terms of the free particle propagator (see also appendix \ref{appPI}), $K_0$, as

\begin{equation}  \label{deltas}
\mathcal{I}_{n}(\{z_{i}\}, \{\tau_{i}\}) =  
    K_0 (x,z_{\Pi(1)};\tau_{\Pi(1)})  K_0 (z_{\Pi(1)},z_{\Pi(2)} ;\tau_{\Pi(2)}-\tau_{\Pi(1)}) \cdots K_0 (z_{\Pi(n)}, y ;T-\tau_{\Pi(n)})\, .
\end{equation}
Here $\Pi$ is the time-ordering permutation of subindices defined so that $0 \leq \tau_{\Pi(1)} \leq  \tau_{\Pi(2)} \leq \cdots \tau_{\Pi(n)} \leq T$, and in $D$ spatial dimensions the free propagator is
\begin{equation} \label{K0inD}
    K_0 (x,y;\tau) = \frac{1}{(4\pi \tau)^{D/2}}e^{-(x-y)^2/4\tau} \theta(\tau)   \,.
\end{equation}

The physical meaning of $\mathcal{I}_{n}$ is manifest: it represents the {\it relative} amplitude\footnote{A {\it relative} amplitude $\Psi$ is an amplitude density per unit {\bf Vol}$^{n+1}$ normalised so that $\int d^D z_1 \cdots d^Dz_n d^Dy \,\Psi =1$. We refer to these contributions as amplitudes even though they are actually the Euclidean version (probabilities) of a true quantum mechanical amplitude.} of propagation starting off from $x$ at time $\tau =0$ and arriving at $y$ at time $\tau = T$ going through the prescribed intermediary points $z_1,z_2,\cdots,z_n$ at their assigned times $\tau_1, \tau_2 ,\cdots, \tau_n$ in time-ordered fashion.

For the potential $V(x)$ with support on $S$ the full series (\ref{fullK}) can be found from (\ref{deltas}) 

\begin{equation} \label{Kper}
K(y,x;T) = \sum_{n=0}^{\infty} \int_S d\sigma_n \cdots \int_S d\sigma_1 \frac{(-\lambda)^n}{n!} \int_0^T d\tau_n \cdots  \int_0^T d\tau_1   \,\mathcal{I}_{n}(\{z_{i}\}, \{\tau_{i}\})\, \theta (T).
\end{equation}
 Notice that the same method can be used for a more general potential by rewriting $V(x) = \int V(z) \delta (x-z) d^D z $ and doing the appropriate replacements, we discuss this matter in section \ref{Vpot}. This is all well known and constitutes one approach to arriving at standard perturbation theory in quantum mechanics \cite{feynman2010quantum}, and it is at this point that we will introduce a series of new developments aimed at treating this problem in a distinct manner.

First we point out that the integrations in (\ref{Kper}) contain all possible time orderings of the variables $\tau_i$, but thanks to permutation symmetry each ordering yields the same contribution after integrating the $z_i$ over $S$. This allows us to fix an arbitrary ordering and cancel the $1/n!$ prefactor (as we do below in (\ref{Khitseries})). For this reason it is now convenient to choose a canonical ordering $\tau_0 \leq \tau_1 \leq \tau_2 \leq \cdots \leq \tau_n\leq \tau_{n+1}:=T$ and define, in accordance with (\ref{eqHPIdef}), the function
\begin{equation} \label{HitTime}
\H_0 (z_1,z_2,\cdots,z_n|y,x;T) := 
\int_0^T d\tau_{n}\int_0^{\tau_n} d\tau_{n-1}\cdots  \int_0^{\tau_2} d\tau_1 \, K_0 (x,z_{1};\tau_{1})K_0 (z_{1},z_{2}; \tau_{2}-\tau_{1}) \cdots K_0 (z_n,y; \tau_{n+1}-\tau_{n})\,,
\end{equation}
that we shall call the $n$-hit function -- it relates to the original hit function (\ref{1hitbar}) through  $\H_0 (z|y,x;T) = T \widebar{\mathcal{H}}_0(z \I y, x; T)$, which illustrates that in this article we have chosen a different normalisation convention than in \cite{PvHz}. For notational convenience we also drop the bar used there to distinguish between normalised and un-normalised probability distributions. The path integral representation of this function is nothing other than (\ref{eqHPIdef}) with the potential $V(x) = 0$ (indicated in (\ref{HitTime}) by the subscript 0)\footnote{Of course we could define an analogous hit function for a non-zero background potential (aside from the localised potential $V$ on $S$) by replacing $K_{0} \longrightarrow K_{U}$, the kernel corresponding to the quantum mechanical system with potential $U(x)$, in (\ref{HitTime}).}.

The $n$-hit function can be normalised into a relative amplitude\footnote{The actual relative amplitude describing the free propagation is the sum over interfering alternatives, recalling that (\ref{HitTime}) is defined with a canonical ordering, which corresponds to the sum over all $n!$ permutations/orderings: $ \frac{1}{T^n}\sum_{\Pi}\H_0 (z_{\Pi(1)},z_{\Pi(2)},\cdots,z_{\Pi(n)}|y,x;T)$.} by noticing that 
\begin{equation}\label{RelNormalisation}
 \frac{n!}{T^n}  \int d^D z_1 \cdots  d^D z_n \,\H_0 (z_1,\cdots,z_n|y,x;T) = K_0 (y,x;T)\,, 
\end{equation}
which also provides the inverse integral transform that recovers the original propagator from the $n$-hit function. In this manner it is consistent to take the $0$-hit function to be the free space propagator, $K_{0}(y,x; T)$. 

Through direct substitution we can now write the propagator (\ref{Kper}) in a perturbative series that involves successive hit functions integrated over $S$
\begin{equation} \label{Khitseries}
    K(y,x;T) = K_0(y,x;T)  - \lambda \int_S d\sigma_1\, \H_0 (z_1| y,x ; T) + \lambda^2 \int_S \int_S d\sigma_1 d\sigma_2 \, \H_0 (z_1,z_2| y,x ; T)+ \cdots 
\end{equation}
This power series in $\lambda$ constitutes the basis of our approach to calculate the propagator and corresponds to counting contributions scattered by the potential exactly once, twice, and so forth. This representation makes clear that knowledge of just the free hit functions is sufficient to recover the kernel in the presence of a localised potential

In dimension $D$ odd, we will show that it is possible to evaluate $\H_0$ in closed form for an arbitrary number of intermediate points $n$. For the case of even $D$, although it is straightforward to calculate the hit function using the various relations we will outline below, a closed formula of similar simplicity has proven elusive so far, except for the special cases implied by (\ref{Hit2D}). These and other results regarding the $n$-hit function will be postponed to a later section, since instead we will focus first on the case $D=3$ of special physical significance. 

\section{Pad\'e representation of the propagator} \label{PadeRepn}

The problem of calculating the propagator in quantum mechanics, or equivalently the heat kernel, in a bounded region of Euclidean space and on compact manifolds has been investigated for a long time and consequently an arsenal of methods and remarkable results are already known \cite{vassilevich2003heat}. With the exception of systems with a certain degree of symmetry the propagator is not known in analytic form and therefore approximations or numerical calculations must be made. A solution of the Dirichlet problem in terms of analytic expressions will be relevant in light of the importance of heat kernel methods used in quantum field theory \cite{Bastianelli:2007jr, Vinas:2010ix, pisanietal}. In this section we propose an analytical solution and claim that this constitutes a new representation for the propagator in bounded regions, or equivalently an analytical expression for the path integral computed  from trajectories that do not intersect a given boundary. We lend support to this claim with numerical evidence in later sections. 

\subsection{Dirichlet boundary conditions}
We are interested in the solution of the Dirichlet problem, that is the calculation of the propagator subject to the amplitudes vanishing on the boundary $S$. This condition can be imposed in the propagator by taking the limit $\lambda \to +\infty$ that makes the potential barrier $S$ impenetrable. It is therefore necessary to determine the limiting behaviour as $\lambda \rightarrow +\infty$ of the perturbative series representation of the propagator in (\ref{Khitseries}), which naturally falls well outside the radius of convergence of the series. C. Grosche has demonstrated that in the presence of a Dirac delta potential in
$N$ points the energy representation of the quantum mechanical propagator is exactly a rational function of the coupling strength $\lambda$ in dimensions $D=1,2,3$ \cite{grosche1990path, grosche1995delta,grosche1993delta}. We propose that while the exact propagator will not (usually) be a rational function for a delta potential with support on a surface $S$, the propagator can nevertheless be approximated by a sequence of rational functions of $\lambda$ to any desired degree of precision. Following this notion we will turn to Pad\'e approximants and use an idea introduced by  C. Bender and S. Boettcher in \cite{bender1994determination} whereby the value of a function at infinity can be obtained under favourable circumstances from the coefficients of its perturbative series using diagonal Pad\'e approximants. 

As a result, our proposed method is to determine a sequence of diagonal Pad\'e approximants, $P_{N}^{N}$ (see appendix \ref{PShApp}), to the series representation of the propagator based on the iterated integrals of the hit function and then to extrapolate to the strong coupling limit, or $N\rightarrow \infty$. As we show in the appendix, these can be expressed as a quotient of determinants that give a rational approximation to the propagator. Following this approach, after calculating the Pad\'e approximant to (\ref{Khitseries}) we are led to conjecture that the difference between the propagator and the propagator in free space is given by the limit of a quotient of determinants 

\begin{equation} \label{Kpade}
\hspace{-2em}K(y,x;T)- K_0 (y,x;T) = \lim_{N \to \infty} 
\frac{ \left| \begin{array}{ccccc}
     0  & c_0 & c_1 & \cdots &c_{N-1}  \\
     c_0& c_1 & c_2 & \cdots & c_N \\
     c_1 & c_2 &\cdots & \cdots & \cdots \\
     \vdots & \vdots & \vdots & \vdots & \vdots \\
     c_{N-1} & \cdots & \cdots & \cdots & c_{2N-1}
\end{array} \right|
}{\left| \begin{array}{cccc}
      c_1 & c_2 & \cdots &c_{N}  \\
     c_2 & c_3 & \cdots  & c_{N+1} \\
     \vdots &  \vdots & \vdots & \vdots \\
     c_{N} & c_{N+1}  & \cdots & c_{2N-1}
\end{array} \right|}   
\end{equation}

\noindent where
\begin{equation} \label{cPade}
    c_n := (-1)^{n+1} \int_S d\sigma_1 \cdots \int_S d\sigma_{n+1} 
  \,  \H_0 (z_1,\cdots,z_{n+1}|y,x;T)\,.
\end{equation}
While we have found strong support for this claim in our numerical calculations, we do note that it remains a conjecture at this point. However it is important to emphasise the significance of (\ref{Kpade})-(\ref{cPade}) as they state how the effect of the boundary on the propagator can be accounted for purely in terms of multiple scatterings at the boundary and that it is calculated explicitly by integrating hit functions over that boundary. To the best of the authors' knowledge this is a novel way to calculate a path integral in a bounded space and here we intend to demonstrate its usefulness. Beyond the utility of a numerical approximation the more formal question of convergence in the $N\to \infty$ limit merits certain interest on its own right, which we hope to investigate in future work. Indeed the strength of the contribution of each $n$-th order scattering at the boundary is quantified by $c_{n-1}$, the conjecture states that the propagation amplitude is the infinite size limit of the quotient of two amplitudes, each one expressed as a determinant. The physical meaning of these individual amplitudes is connected to the multiple scatterings at the boundary in a combinatorial manner and a clearer understanding of the nature of these individual amplitudes might pave the way for an actual proof of the conjecture.

Let us also draw attention to the fact that, although completely different in content, (\ref{Kpade}) resembles in spirit the original formulation of the Gelfand-Yaglom theorem \cite{gelfand} for a Schr\"odinger problem with Dirichlet boundary conditions in which the effect of a potential on the path integral normalisation is expressed as a finite quotient of functional determinants that were calculated through a limiting process from finite dimensional matrix determinants.

Together (\ref{Kpade}) and (\ref{cPade}) state one of the principal results of the present work, whilst simultaneously motivating us to determine the analytic form of the free $n$-hit functions for their practical use in (\ref{Kpade}). It is also clear from a practical point of view that a finite $N$ must be chosen at some point to carry out an actual approximate calculation of the propagator. In this sense our result also constitutes a pragmatic approximation that in contrast to conventional series representations should get more accurate the stronger the coupling.

\subsection{Other static potentials} \label{Vpot}

Before continuing, we pause to note that an arbitrary potential $V(x)$ may not be tractable with the present approach as the sequence of Pad\'e approximants may not converge, yet we wish to illustrate that a wide family of potentials could still be treated with this method. We start off by writing the series expansion (\ref{fullK}) for the potential
\begin{equation}
  V_{\lambda} (x) = \lambda \int d^D z\, V(z) \delta (x-z) \,.
\end{equation}
Following the procedure outlined is tantamount to the formal substitution $\int_S \,d\sigma_i \to \int d^Dz_i\, V(z_i)$ everywhere. In this case the parameter $\lambda$ could be an appropriate coupling extracted from the physical potential or could be artificially introduced; in the latter case after calculating the perturbative series analogue to (\ref{Khitseries}) the limit $\lambda \to 1$ is taken. The following diagonal Pad\'e representation for the propagator is obtained

\begin{equation} \label{KVPade}
  \hspace{-2em}  K(y,x;T)  = \lim_{N \to \infty} 
 \frac{ \left| \begin{array}{cccc}
         \varphi_0 &  \varphi_1 &\cdots & \varphi_N  \\
        C_{1}& C_{2} &\cdots & C_{N+1} \\
        \vdots & \vdots & \vdots & \vdots \\
        C_N & C_{N+1} & \cdots & C_{2N}
   \end{array} \right| }{\left| \begin{array}{cccc}
        1  & 1 &\cdots & 1  \\
        C_{1}& C_{2} &\cdots & C_{N+1} \\
        \vdots & \vdots & \vdots & \vdots \\
        C_N & C_{N+1} & \cdots & C_{2N}
   \end{array} \right|} 
\end{equation}

\noindent where $\varphi_L = \sum_{p=0}^L C_p$ and
\begin{equation}
    C_0 = K_0(y,x;T), \qquad C_n = (-1)^n 
    \int d^D z_1 \cdots \int d^D z_n\, V(z_1)\cdots V(z_n)\H_0 (z_1,\cdots,z_n|y,x;T) ,\quad n\geq 1\,.
\end{equation}
Note that, as we shall show below, in contrast to standard perturbation theory, here there will be no requirement that the expansion parameter, $\lambda$, be small.

 In contrast to (\ref{Kpade}) we may now also consider the non-diagonal Pad\'e  approximants: from the numerical standpoint this is often useful to speed up convergence as they provide more information on the convergence behaviour of the sequence, in which case the analogue to (\ref{KVPade}) can be easily obtained from (\ref{PadeMN}) by taking $M\neq N$. It will be interesting to investigate the kind of potentials that this method can be applied to. Although the results presented below lend weight to the argument that this approach will be well suited to localised potentials, the authors venture that it in principle may be applicable as a new series representation of the kernel for some more general potentials as well. It is hoped that further investigation will establish how widely this method can be used.

\section{Determining the $n-$hit Function}
Based  on the proposal outlined above, in this section we calculate the $n$-hit function in the special case of dimension $D = 3$ and then give a general formula, (\ref{HitBromwich}), in arbitrary dimension. From these results we derive relations between the $n$-hit functions for differing $D$ and $n$ which allow us to transform between different values of these parameters with simple operations on the functions. 
\subsection{3-dimensional hit function}

It turns out that $D = 3$ dimensions is a special case for which the iterated integrals in (\ref{HitTime}) can be calculated straightforwardly. Using the relations in appendix \ref{apH3D}, the key observation is that the function $f_{\alpha} (\tau) = \frac{1}{\tau^{3/2}}\e^{-\alpha^2 /\tau}$ is a reproducing kernel over $\tau$ in the sense that 
\[
\int_{0}^{T} f_{\alpha} (\tau-\tau') f_{\beta} (\tau') ~d\tau' = \sqrt{\pi} \frac{\alpha + \beta}{\alpha \beta} f_{\alpha + \beta} (\tau)\,,
\]
so that in terms of the (finite-interval) convolution $[f_{\alpha}\ast f_{\beta}](\tau) = \sqrt{\pi} \frac{\alpha + \beta}{\alpha \beta} f_{\alpha + \beta} (\tau)$. This result allows us to perform all integrals recursively in (\ref{HitTime}) to determine the $n$-hit function for arbitrary order\footnote{It is pleasant to note that this property is the time-analogue of the reproducing kernel in space for the same function:

\[
\int_{\mathbb{R}^3} f_{|y-z|} (T-\tau) f_{|z-x|}(\tau) ~d^3 z = \pi^{3/2} f_{|y-x|}(T) \]
Here position space vectors are added; over time it is their lengths that are to be added. Both behaviours can be traced back to the fact that free propagation  forgets about the prior points it has previously passed through and is isotropic so that the amplitude depends not on the relative orientations of the edges of the path between the points $z_{i}$, but rather on their lengths only.}.

Repeated application of the convolution property and the results for the integrals given in appendix \ref{apH3D} yield a closed formula for the $n$-hit function in $D = 3$:
\begin{equation} \label{Hit3D}
\H_0 (z_1,\cdots, z_n |y,x;T) = \frac{1}{(4\pi)^{\frac{2n+3}{2}}} \frac{e^{- \frac{\Delta^2}{4T}  }}{T^{3/2}} \frac{\Delta}{\Delta_1 \Delta_2 \cdots \Delta_{n+1}}\,\theta (T)\,,
\end{equation}
where we introduced $\Delta_k : = |z_{k} - z_{k-1}|$ and defined $z_{0} := x$ and $z_{n+1}:=y$, so that $\Delta=\sum_{i=1}^{n+1} \Delta_i$ is the total length of the polygonal path from $x$ to $y$ with prescribed vertices $\{z_i\}$. We shall make frequent use of this notation for the remainder of this article. 

To illustrate this result we give the explicit form of the first two hit functions as
\begin{eqnarray}
\H_0 (z|y,x;T) & = & \frac{1}{(4\pi)^{5/2}T^{\frac{3}{2}}} \left(\frac{|y-z| + |z-x|}{|y-z| |z-x|} \right) \e^{ - \frac{1}{4T}\left(|y-z| + |z-x| \right)^2 } \,\theta (T),\\
\H_0 (z_1,z_2 | y,x ; T) &=& \frac{1}{(4\pi)^{7/2}T^{\frac{3}{2}}} 
\left( \frac{ |x-z_1| + |z_1 -z_2| + |z_2 -y| }{ |x-z_1||z_1-z_2||z_2-y| }\right)
\e^{- \frac{1}{4T}\left(|y-z_2| + |z_2 -z_1| + |x-z_1| \right)^2 }\,\theta (T)\,.
\label{H2D1Id}
\end{eqnarray}
We remark here about the normalisation of these functions in comparison to that used in previous work. In \cite{PvHz} the hit function was normalised such that integrating it over space gave unity (a second, unnormalised distribution, denoted $\widebar{\mathcal{H}}$, was also introduced). Here, to convert the $n$-hit functions into correctly normalised probability distributions on the particle trajectories, we would divide $\H_0 (z_1,\cdots, z_n |y,x;T)$ by $ K_{0}(y, x; T)T^{n} / n!$ such that the multiple integral $\frac{ n!} { T^{n}K_{0}(y, x; T)} \prod_{i=1}^{n} \int d^{D}z_{i} \, \H_0 (z_1,\cdots, z_n |y,x;T) = 1$ (c.f. (\ref{RelNormalisation})). 

\subsection{Arbitrary $n$-hit function in dimension $D$}

In the general case the evaluation of the hit function defined in (\ref{HitTime}) is greatly facilitated by using the ``energy representation'' of the kernel, its Fourier transform with respect to transition time $T$,
\begin{equation}
\widehat{K} (y,x;\omega) = \int_{-\infty}^{\infty} d T~ K (y,x; T)  \e^{i \omega  T}\,.
\label{eqKEnergy}
\end{equation}
Here we use this representation of the kernel to manipulate the integral representation of the $n$-hit function into the form of an inverse Mellin transform.

\begin{figure}
\centering
  \includegraphics[width=0.35\textwidth]{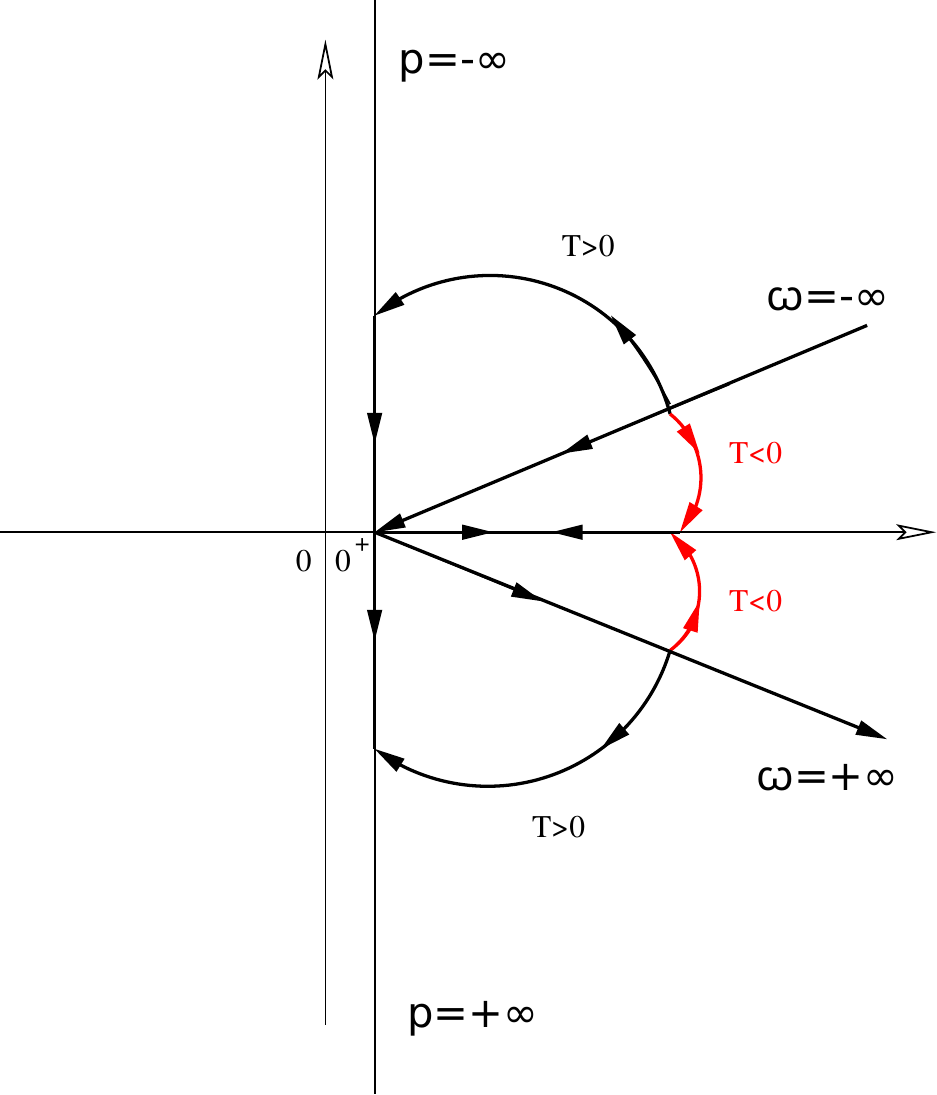}
  \caption{The complex plane illustrating the deformation of the integration contours that leads to the change of variables between $\zeta$ and $p$.}
  \label{Contour}
\end{figure}

We begin with the observation that the $n$-hit  function can be written as (in fact this result holds for the hit function defined with respect to an arbitrary background potential by replacing $K_0 \to K_U$, cf. footnote in Sec. \ref{QuantP})

\begin{equation} \label{Hitdelta}
 {\H}_{0} (z_1, z_2, \cdots, z_n|y,x;T) = 
\prod_{i=1}^{n+1}\int_{-\infty}^{\infty}\!\! dt_i \, \delta \Big( \sum_{k=1}^{n+1} t_k -T \Big) K_0 (y,z_{n}; t_{n+1})
K_0 (z_{n},z_{n-1}; t_n) \cdots 
K_0 (z_{1},x; t_{1}) \,,
\end{equation}
 where we made use of the change of variables $t_i:= \tau_i - \tau_{i-1}$ for $1 \leq i \leq n+1$, again with $\tau_0 :=0$ and $\tau_{n+1}:=T$, so that the $t_i$ now represent the lengths of proper-time intervals. Using now the well known Fourier representation of the Dirac delta distribution $\delta (t) =  \frac{1}{2\pi} \int_{-\infty}^{\infty}\!\! d\omega~ \e^{i \omega t }$ the convolution theorem yields the integral representation (again, replacing the free propagators on the right-hand side by their counterparts corresponding to some other background potential yields an analogous integral representation of the $n$-hit function for that potential)
\begin{equation} \label{hitomega}
 {\H}_0 (z_1, z_2, \cdots, z_n|y,x;T) = \int_{-\infty}^{\infty} \frac{d\omega}{2\pi} e^{-i\omega T}  \widehat{K}_0 (y,z_n;\omega)\cdots \widehat{K}_{0}(z_2,z_1;\omega) \widehat{K}_0 (z_1,x;\omega) \,.
\end{equation}
A physical interpretation of (\ref{hitomega}) is that as the particle is scattered multiple times in vacuum (or in a static potential) the amplitudes corresponding to a given fixed energy are multiplied, that is the scatterings happen independently and do not change the energy; we then calculate the total amplitude by integrating over the values of energy at which this scattering took place. In our conventions, (\ref{hitomega}) contains the Fourier transform of the free propagator according to (\ref{eqKEnergy}) which for the free particle is known\footnote{The relevant identity is (see, for example, \cite{gradshteyn2007table}) \[
\int_0^{\infty} dt t^{\nu-1} \e^{-i\gamma t + i\frac{\beta}{t} } = 2 \left(\frac{\beta}{\gamma} \right)^{\nu/2} \e^{-i\nu \pi/2} K_{-\nu}(2 \sqrt{\beta \gamma})\qquad  \Im (\gamma) \leq 0,\quad ~\Im (\beta) \geq 0
\]},  for dimension $D$, in terms of the Macdonald function (modified Bessel function of the second kind), $K_{\frac{D-2}{2}}$ (see \cite{KleinertBook}, for example). Indeed, with $\nu := \frac{2-D}{2}$ one obtains for arbitrary $D$,
\begin{equation}
  \widehat{K}_0 (y,x;\omega) =  \frac{2}{(4\pi)^{D/2}} \left(\frac{|x-y|}{2\sqrt{-i\omega}} \right)^\nu   K_{-\nu}(\sqrt{-i\omega } |x-y|)  \, .
 \end{equation}
Substituting this into (\ref{hitomega}) into we arrive at the following Fourier integral representation
\begin{equation} \label{HitFourier}
  {\H}_0 (z_1,\cdots, z_n|y,x;T) = \mathcal{C}_{D,n} \int_{-\infty}^{\infty} \frac{d\omega}{2\pi} \frac{\e^{-i\omega T}}{ (-i\omega )^{\frac{(n+1)\nu}{2}}} K_{-\nu}(\sqrt{-i\omega }|x-z_1|)\cdots K_{-\nu} (\sqrt{-i\omega } |z_n-y|)\, .
\end{equation} 
where we introduced the prefactor
\[
\mathcal{C}_{D,n}= \left(2\pi \right)^{-D(n+1)/2} (|x-z_1|\cdots |z_n-y|)^{\nu}\, .
\]
We now introduce the complex variable $\zeta =\sqrt{- i\omega }$ and proceed to deform the contour of integration using Cauchy's theorem; when $T<0$ we can collapse the path onto the real axis on the complex $\zeta$ plane, leading to the vanishing of the integral, as required by the Heaviside $\theta$ functions in (\ref{K0inD}), whereas in the case $T>0$ the contour can no longer be deformed in this manner but rather can be deformed to run parallel to the imaginary axis shifted a small amount $0^+$ towards the right in accordance with the Feynman prescription -- see Fig. \ref{Contour}. This straight path can be parameterised in terms of a new variable $p \in (-\infty + i0^+ ,  \infty+i0^+)$ and the natural change of variables in the integrand is $p=-i\sqrt{- i\omega }$ so that this variable is the Wick-rotated momentum corresponding to energy $\omega$ -- the resulting contour in the complex plane parameterised by $p$ is shown in Fig. \ref{Contour}. With these considerations we arrive at the integral representation\footnote{To make manifest that this integral representation is real one can take advantage of the relation for $x\in \mathbb{R}$: $K_{-\nu}(-ix)= \frac{\pi i}{2}(-1)^\nu H_{\nu}^{(1)} (x)$.}

\begin{equation} \label{HitBromwich}
     {\H}_0 (z_1,\cdots, z_n|y,x;T) = 2\mathcal{C}_{D,n} \int_{i0^+ -\infty}^{i0^+ + \infty} \frac{dp}{2\pi i} \frac{p\, \e^{- p^2 T}}{(-ip)^{(n+1)\nu}} K_{-\nu}(-ip |x-z_1|)\cdots K_{-\nu}(-ip |y-z_n|)\, \theta (T) \,.
\end{equation}
Notice that this is in fact the Bromwich integral for the inverse Laplace transformation, or equivalently the Mellin transform of the integrand. A significant advantage of the formulae (\ref{HitFourier}) and (\ref{HitBromwich}) over (\ref{HitTime}) or (\ref{Hitdelta}) is that the $n$ or $n+1$ integrations required to calculate the $n$-hit function are ultimately reduced to only one; this is an improvement if such functions are to be calculated numerically (where there are already well developed algorithms for estimating such integrals). In the next sections we shall use this integral representation to obtain various functional identities between hit functions and then show how it can be evaluated to determine specific hit functions in different dimensions and of different orders.

\section{Identities amongst $n$-hit functions}
\label{secIds}
It turns out that there are various identities relating the generalised hit functions for different spatial dimension or number of intermediate points, $n$. In this section, for clarity we will write the space dimension $D$ explicitly in the $n$-hit function by adding a super-index in the form $\H_0^{(D)}$.

\subsection{Changing the order $n$ and dimension $D$}
We begin by considering hit functions defined in different dimension. By writing out (\ref{HitTime}) explicitly,
\begin{equation}\label{eqHintTime} 
    \H_0^{(D)}(z_1,\cdots,z_n|y,x;T) = \frac{1}{(4\pi)^{(n+1)D/2}}\int_0^T d\tau_n \cdots \int_0^{\tau_2} d\tau_1 \,
   \prod_{i=1}^{n+1} \frac{\e^{- \Delta_i^2/4(\tau_i - \tau_{i-1})}}{(\tau_i - \tau_{i-1})^{D/2}}\,\theta(T)\, ,
\end{equation}
we see that the dependence on the dimension appears explicitly in the denominator, the prefactor and, implicitly in the definition of each $\Delta_{i} = |z_i - z_{i-1}|$ as a distance in $D$-dimensional Euclidean space. As a direct consequence of differentiation under the integral the following functional relation holds
\begin{align} \label{Duptwo}
\H_0^{(D+2)}(z_1,\cdots,z_n|y,x;T)&=  \prod_{i=1}^{n+1} \left(\frac{-1}{2\pi}\frac{1}{\Delta_i}\frac{\partial}{\partial \Delta_i} \right)  \H_0^{(D)}(z_1,\cdots,z_n|y,x;T) \\
&:= \mathcal{D}^{+}_n\H_0^{(D)}(z_1,\cdots,z_n|y,x;T)\, ,
\end{align}
where we denote the operation of raising the dimension by $\mathcal{D}^{+}_n$ that acts on a hit function of order $n$ (see below for its lowering partner). Here it is important to stress that (\ref{Duptwo}) must be interpreted as a functional relation and not an equality; in particular the meaning of the $\Delta_i$ is different on both sides as they are to be calculated in spaces of dimensions $D+2$ and $D$ respectively. Nevertheless this identity allows one to calculate hit functions in higher dimensions of the same parity once a given $n$-hit function is explicitly known. To illustrate this, specific examples are given in section (\ref{examplesB}). It is also possible to write the inverse functional relation that will lower the dimension from $D\to D-2$ by inverting, with the appropriate boundary conditions at $\Delta_i\to \infty$, the operation defined in (\ref{Duptwo}):
\begin{align} \label{Ddowntwo}
    \H_0^{(D-2)}(z_1,\cdots,z_n|y,x;T) &= \prod_{i=1}^{n+1}\left( 2\pi \int^{\infty}_{\Delta_i} d\Delta_i \, \Delta_i \right) \H_0^{(D)}(z_1,\cdots,z_n|y,x;T) \\
    &:= \mathcal{D}^-_n \H_0^{(D)}(z_1,\cdots,z_n|y,x;T)\,,
\end{align}
which defines the dimension-lowering operator $\mathcal{D}_n^-$ of order $n$ (of course in the form we have written these operations it is clear that $\mathcal{D}_{n}^\pm$ depend upon the intermediate points but the meaning of the abstract operators acting on the hit functions should be clear).

Turning now to relations between hit functions of different order, from (\ref{HitTime}) one immediately sees how to incorporate an additional intermediate point, $z_n$ say, into the hit function which allows us to define an order-raising operator $\mathcal{N}^+_D(z_n)$, that will insert the additional point $z_{n}$ in the $D$-dimensional hit function according to\footnote{The same identity holds if a potential $U$ is present by replacing $\H_0 \to \H$ and $K_0 \to K$, cf. footnote in section \ref{QuantP}.} 
\begin{align} \label{convol}
    \H_0^{(D)} (z_1,\cdots, z_n|y,x;T) &= \int_0^T d\tau \, \H_0^{(D)} (z_1,\cdots,z_{n-1}|z_n, x;\tau) 
    K_0(y,z_n;T-\tau) \\
    &:= \mathcal{N}^+_D(z_n) \H_0^{(D)} (z_1,\cdots,z_{n-1}| y, x;\tau) \,.
\end{align}
This can also be expressed as a convolution product $[\H_0 (z_1,\cdots,z_{n-1}|z_n, x)\ast K_0(y,z_n)](T)$ which produces the required insertion point using the appropriate kernel. We can invert (\ref{convol}) by introducing a  Dirac $\delta$-distribution under the integral with an operator, $\mathcal{N}_D^-(z_n)$, that will lower the order of the hit function by removing the point $z_n$ by integrating it out (in $D$-dimensions)
\begin{align} \label{convolinv}
   \H_0^{(D)} (z_1,\cdots, z_{n-1}|y,x;T) &= \frac{1}{2} \int d^D z_n \, \Big(  \frac{\partial}{\partial T} - \nabla_y^2 \Big)  \H_0^{(D)} ( z_1,\cdots, z_{n}|y,x;T)\\
    &:=    \mathcal{N}^-_D(z_n) \H_0^{(D)} (z_1,\cdots,z_{n}| y, x;T) \,,
\end{align}
where $\nabla_y^2$ is the Laplacian acting only on the $y$ coordinates. In fact this follows from the Schr\"odinger-like equation that the hit function satisfies that is inherited from the Schr\"odinger equation, a generalisation of the Green function equation satisfied by the kernel itself:
\begin{equation}\label{eqHGreen}
    \half  \Big(  \frac{\partial}{\partial T} - \nabla_{y}^2 \Big)  \H_0^{(D)} (z_1,\cdots, z_{n}|y,x;T) = \delta^D(z_{n}-y)\H_0^{(D)} (z_1,\cdots, z_{n-1}|y,x;T)\,.
\end{equation}

Consider now the special case $D=1$: due to its exceptional simplicity there is another manner to relate the $n$ to $n+1$-hit functions. To this end observe that for $D=1$ (\ref{HitBromwich}) becomes

\begin{equation}\label{eqHitD1}
\H^{(1)}_{0}(z_1, \cdots, z_n|y,x;T) = \left(\frac{i}{2}\right)^n  \int_{i0^+ -\infty}^{i0^+ + \infty} 
\frac{dp}{2\pi p^n} e^{-p^2 T + i p \Delta}\, \theta (T)\,.
\end{equation}
Notice that the only dependence on $n$ is in the prefactor and the power of $p$ in the integrand whilst the dimensional dependence only enters implicitly through $\Delta$. By differentiation we obtain

\begin{equation} \label{Hstepn}
  \H^{(1)}_{0}(z_1,\cdots,z_{n-1}|y,x;T)
  =  -2 \frac{\partial}{\partial \Delta} \H^{(1)}_{0}(z_1,\cdots,z_n|y,x;T)\,,
  \end{equation}
which must be interpreted, as above, as a functional relation considering that the meaning of $\Delta$ is different on both sides of (\ref{Hstepn}) on account of the change in the number of intermediate points; the inverse relation follows from considerations similar to those made for (\ref{Ddowntwo})

\begin{equation} \label{Hstepninv}
\H^{(1)}_{0}(z_1,\cdots,z_n|y,x;T) = \half \int_{\Delta}^{\infty} d\Delta \, \H^{(1)}_{0}(z_1,\cdots,z_{n-1}|y,x;T) \,.
\end{equation}
The commutative diagram that follows shows the mapping between the different $n$-hit functions
\begin{figure}[h] 
\centering
    \begin{tikzpicture}[node distance=1.25cm and 2cm]
    \tikzstyle{line} = [draw, {Stealth[scale=1.2]}-{Stealth[scale=1.2]}]
    \tikzstyle{linel} = [draw, dash pattern=on 3pt off 3pt, color=gray, -{Stealth[scale=1.2]}]
     \tikzstyle{blockL} = [minimum width=1em]
    \tikzstyle{blockR} = [minimum width=1em]
    
\node (init) {};
\node [blockL] (DDn) {$\H_0^{(D+2)}(n)$} ;
\node [blockR] [right=of DDn] (DDnn) {$\H_0^{(D+2)}(n+1)$};
\node [blockL] [below=of DDn] (Dn) {$ \H_0^{(D)}(n)$} ;
\node [blockR] [right=of Dn, below=of DDnn] (Dnn) {$\H_0^{(D)}(n+1)$};

\node [blockL] [left=0.75cm of DDn] (DDnl) {};
\node [blockR] [right=0.75cm of DDnn] (DDnnr) {};
\node [blockL] [left=0.75cm of Dn] (Dnl) {};
\node [blockL] [right=0.75cm of Dnn] (Dnnr) {};

\node [blockL] [above=0.75cm of DDn] (DDnu) {};
\node [blockR] [above=0.75cm of DDnn] (DDnnu) {};
\node [blockL] [below=0.75cm of Dn] (Dnd) {};
\node [blockL] [below=0.75cm of Dnn] (Dnnd) {};

\path [line] (DDn) -- node [midway, above=0.25em] {\scriptsize $\mathcal{N}_{D+2}^{\pm}$} (DDnn);
\path [line] (DDn) -- node[midway, left=0.25em] {\scriptsize $\mathcal{D}_{n}^{\pm}$} (Dn);
\path [line] (DDnn) -- node[midway, right=0.25em] {\scriptsize $\mathcal{D}_{n+1}^{\pm}$} (Dnn);
\path [line] (Dn) --  node [midway, below=0.2em] {\scriptsize $\mathcal{N}_{D}^{\pm}$} (Dnn);

\path [linel] (DDnl) -- (DDn);
\path [linel] (DDnnr) -- (DDnn);
\path [linel] (Dnl) -- (Dn);
\path [linel] (Dnnr) -- (Dnn);

\path [linel] (DDnu) -- (DDn);
\path [linel] (DDnnu) -- (DDnn);
\path [linel] (Dnd) -- (Dn);
\path [linel] (Dnnd) -- (Dnn);

\end{tikzpicture}
\caption{An illustrative part of the infinite commutative diagram connecting the hit functions of different order in different dimension: $\H_0^{(D)}(n)$ denotes the $n$-hit function in dimension $D$
\label{comdiag}
}
\end{figure}
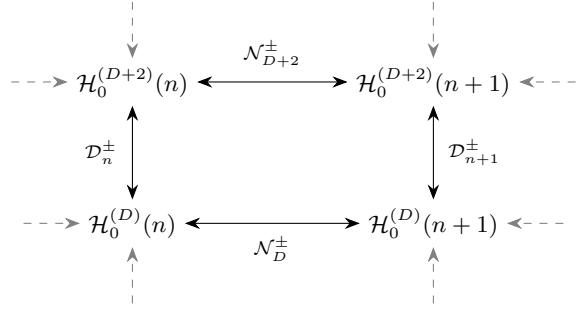

As a consequence of the relations described we draw the conclusion that all $n$-hit functions are related in dimensions with the same parity, and in a given dimension different orders of hit function (values of $n$) are also related, as shown in the diagram in Fig. \ref{comdiag}

The relations shown in the commutative  diagram, and described by  (\ref{Duptwo}),(\ref{Ddowntwo}),(\ref{convol}) and (\ref{convolinv}) are general. In the following section we shall apply them to obtain closed formulae for the odd dimensional case on account of its special simplicity (since all $n$-hit functions can be related to the $D=1$ or $D=3$ case). In the even dimensional case we will provide simple integral representations for $n>1$. In the special case with $D=1$ spatial dimensions the additional identities (\ref{Hstepn}) and (\ref{Hstepninv}) hold which help us to construct all hit functions as a basis for the odd dimensional case. We also provide some results in even dimensions (in particular $D=2$).

\subsection{Examples} \label{examplesB}
Let us illustrate the use of the integral representation (\ref{HitBromwich}) and the relations derived in this section to determine the $n$-hit function for some cases of interest. In particular we shall focus on the results for $D = 1$ and $D = 3$ and find the hit function for arbitrary $n$.

First, to recover the result (\ref{Hit3D}), we use $\nu = -\frac{1}{2}$ and the well-known analytic expression for the Bessel functions, $K_{1/2}(x) = \sqrt{\frac{\pi}{2}} \frac{\e^{-x}}{\sqrt{x}}$. Thus in $D =3$ the integral representation, (\ref{HitBromwich}), of the $n$-hit function turns into
\begin{equation}
     \H_0^{(3)}(z_1,\cdots,z_n|y,x;T) = \frac{2}{(4\pi)^{n+1}} \Big( \prod_{i=1}^{n+1} \Delta_{i} \Big)^{-1} \int_{-\infty}^{\infty} \frac{dp}{2 \pi i} \, p\, \e^{-Tp^{2} + ip \sum_{i = 1}^{n+1} \Delta_{i} } \theta(T)\, .
\end{equation}
Completing the square in the exponent and shifting integration variables results in a Gaussian integral that evaluates immediately to (\ref{Hit3D}). We also note that this result is, of course, compatible with the convolution identity, (\ref{convol}). Indeed, using (\ref{int2}), we immediately verify
\begin{align}
  \mathcal{N}_{3}^{+}(z_{n}) \H_0^{(3)}(z_1,\cdots,z_{n-1}|z_{n},x;\tau) &=  \int_{0}^{T} d\tau \, \theta (T-\tau) \frac{\e^{-\frac{(z_{n}-y)^{2}}{4 (T-\tau)}}}{(4\pi (T-\tau))^{\frac{3}{2}}}  \H_0^{(3)}(z_1,\cdots,z_{n-1}|z_{n},x;\tau)\\ &=   \frac{1}{(4\pi)^{\frac{2n+1}{2}}}  \frac{\Delta - \Delta_{n+1}  }{\Delta_1 \Delta_2 \cdots \Delta_{n}}    \int_{0}^{T} d\tau \, \frac{\e^{-\frac{\Delta_{n+1}^2}{4 (T-\tau)}}}{(4\pi (T-\tau))^{\frac{3}{2}}}  
  \frac{e^{- \frac{(\Delta -\Delta_{n+1})^2 }{4\tau}  }}{\tau^{3/2}} \,\theta (T)\nonumber \\
    &= \frac{1}{(4\pi)^{\frac{2n+3}{2}}}  \frac{\Delta}{\Delta_1 \Delta_2 \cdots \Delta_{n+1}}  \frac{\e^{-\frac{\Delta^{2}}{4T}}}{T^{\frac{3}{2}}} \,\theta (T)\\
    &= \H_0^{(3)}(z_1,\cdots,z_{n}|y,x;T)\,.
\end{align}
Likewise, for the case $D=1$ it is also possible, with a little more work, to arrive at a closed form result for the hit function of arbitrary order. We can use this to verify the formulas \eqref{Duptwo} and \eqref{Ddowntwo} for changing dimension. A key relation is \eqref{Hstepninv} which defines a useful recursion relation between hit functions of different order. In this case, the special form of the one hit ($n=1$) function makes it possible to solve this relation in closed form. 

Although we have already quoted the result for $\widebar{\H}_0^{(1)}(z|y,x;T)$ in \eqref{1hitbar} of the introduction, here we take advantage of the opportunity to recalculate it using the new ``master formula'' derived in \eqref{HitBromwich}. Indeed, for $D=1$ this reduces to \eqref{eqHitD1} so with $n=1$ we must calculate
\begin{equation}
    \H^{(1)}(z|y,x;T) = \frac{i}{2}  \int_{i0^+ -\infty}^{i0^+ + \infty}
\frac{dp}{2\pi p} e^{-p^2 T + i p (\Delta_1 + \Delta_2)}\,.
\end{equation}
It is clear that we must take care with the pole at $p=0$ (this is already apparent in \eqref{eqHitD1}). To deal with this we separate the integral as follows (this treatment is equivalent to using the prescription for the pole $\frac{1}{p+i0} = \mathcal{P}(1/p) - i\pi\delta(p)$)

\begin{equation}
{\H}_0 (z|y,x;T) =\frac{1}{2}\Big(-\int_{\gamma(\epsilon)} \frac{dp}{2\pi i} \frac{e^{-p^2 T + i p\Delta }}{p}    - \dashint_{-\infty}^{\infty} \frac{dp}{2\pi } 
\frac{\sin(p\Delta)}{p}  e^{-\half p^2 T  }\Big)\,,
\end{equation}
where $\gamma(\epsilon)$ is the semi-circle of radius $\epsilon$  above the pole at the origin -- the second is a principle value integral that excludes the interval $(-\epsilon, \epsilon)$ and the limit $\epsilon \rightarrow 0^+$ is understood. The first integral is evaluated as half the contribution given by the residue theorem at the pole, while the second is a known integral that evaluates to an error function so we reproduce the prior result quoted in \eqref{1hitbar},
\begin{equation}
{\H}_0 (z|y,x;T) = \frac{1}{4}\bigg(1 - \erf \left(\frac{\Delta}{2\sqrt{T}} \right)\bigg) = \frac{1}{4} \erfc\left( \frac{\Delta}{2\sqrt{T}} \right)\,,
\end{equation}
once we put $m = \frac{1}{2}$ and recall the change in the normalisation of the hit function adopted in this paper. It is now easy to verify \eqref{Duptwo} and \eqref{Ddowntwo} using \eqref{H2D1Id} which involve two integrations and two differentiations respectively. 

Using this result in the recursion relation \eqref{Hstepninv} shows that the one-dimensional $n$-hit function can be expressed in terms of the well-known iterated integrals of the complementary error function (see, for example, \cite{AbSt,DLMF}) defined by 
\begin{align}
    \ierf{n}(z)\, & := \int_{z}^{\infty} dz'\, \ierf{n-1}(z')\\
    \ierf{-1}(z) &= \frac{2}{\sqrt{\pi}}\e^{-z^2}\\
    \ierf{0}(z) &= \textrm{erfc}(z)\,.
\end{align}
With this notation we solve the recursion relation via
\begin{align}\label{eqH0Erfc}
    \H_0^{(1)}(z_1,\cdots,z_{n}|y,x;T) &= \frac{1}{4}\Big(\frac{1}{2} \int_\Delta^\infty d\Delta \Big)^{n-1}\erfc\left( \frac{\Delta}{2\sqrt{T}} \right)\ \nonumber \\
    &= \frac{1}{4}T^{\frac{n-1}{2}}\ierf{n-1}\left( \frac{\Delta}{2\sqrt{T}} \right)\,,
\end{align}
in which, of course, $\Delta$ is interpreted according to the order $n$. Using the integral definition of $\ierf{n}$ given above, and the general result for \eqref{Hit3D} in three dimensions it is a simple exercise in induction to prove that the identities \eqref{Duptwo} and \eqref{Ddowntwo} hold between dimension $D=1$ and $D=3$ for arbitrary order $n$. For completeness we supply the functional form of these results for $n=2$ and $3$:
\begin{align}\label{eqH23D1}
    H_0^{(1)}(z_1, z_{2}|y,x;T) &= \frac{\sqrt{T}}{4} \left[ \frac{\e^{-\frac{\Delta^2}{4T}}}{\sqrt{\pi}} - \frac{\Delta}{2\sqrt{T}} \erfc \left( \frac{\Delta}{2\sqrt{T}} \right)\right]\,,  \\
    H_0^{(1)}(z_1, z_{2}, z_{3}|y,x;T) &= \frac{T}{16}\left[-2\frac{\Delta}{2\sqrt{T}} \frac{ \e^{-\frac{\Delta^2}{4T}}}{\sqrt{\pi}} + \left(2\frac{\Delta^2}{4T} + 1\right) \erfc \left( \frac{\Delta}{2\sqrt{T}} \right)\right]\,, \label{eqH3D1}
\end{align}
where $\Delta$ is to be formed using $2$ and $3$ intermediate points respectively.

Since the dimensional shift formula \eqref{Duptwo} allows us to use the one-dimensional hit functions to produce all hit functions in odd dimension, it is of interest to give more explicit formulas for the former. As illustrated in \eqref{eqH23D1} and \eqref{eqH3D1}, we observe that integrating elements of the set $\{ \erfc(z), \e^{-z^{2}} \}$ produces polynomials in $z$ multiplied by these same elements. Thus this form of closure under integration implies that the hit function in one dimension is built from polynomials in $\frac{\Delta}{2\sqrt{T}}$ multiplied by the error function or Gaussian exponential of the same variable. Combining the recursion relation with the integral representation \eqref{eqHitD1} and employing the generalisation $\frac{1}{(p + i0)^n} = \mathcal{P}(\frac{1}{p^n}) +i \frac{(-1)^n}{(n-1)!}\pi\delta^{(n-1)}(p)$ it is possible to determine the hit function in terms of special functions
\begin{equation} \label{eqH0Hyp} 
     \H_0^{(1)}(z_1,\cdots,z_{n}|y,x;T) = \frac{\e^{-\frac{\Delta^2}{4T}}}{4 \sqrt{\pi}(n-1)!}T^{\frac{n-1}{2}}\left[ \Gamma\Big(\frac{n}{2}\Big) \IFI{\frac{n}{2}}{\frac{1}{2}}{\frac{\Delta^2}{4T}} - (n-1)\frac{\Delta}{2\sqrt{T}}\Gamma\Big(\frac{n-1}{2}\Big)\IFI{\frac{n+1}{2}}{\frac{3}{2}}{\frac{\Delta^2}{4T}}   \right]\,,
\end{equation}
where $\IFI{a}{b}{z}$ is a confluent hypergeometric function of the first kind \cite{AbSt, gradshteyn2007table}. Of more practical use we can relate the hypergeometric functions and parabolic cylindrical functions and write the latter in terms of derivatives of the error function to arrive at the helpful identities
\begin{align}
    \IFI{\frac{n}{2}}{\frac{1}{2}}{z^2} &=(-1)^{n-1}\frac{\Gamma\Big[\frac{n+1}{2}\Big]}{2(n-1)!}\frac{d^{n-1}}{dz^{n-1}} \Big[ \e^{z^{2}} \Big( \erfc(z) + (-1)^{n-1} \erfc(-z) \Big)\Big]\,,   \\
        \IFI{\frac{n+1}{2}}{\frac{3}{2}}{z^2} &=(-1)^{n-1}\frac{\Gamma\Big[\frac{n}{2}\Big]}{4z(n-1)!} \frac{d^{n-1}}{dz^{n-1}} \Big[ -\e^{z^{2}} \Big( \erfc(z) - (-1)^{n-1} \erfc(-z) \Big)\Big]\, .
\end{align}
If we use these in (\ref{eqH0Hyp}) then it simplifies substantially to a derivative representation
\begin{equation}
     {\H}_0^{(1)} (z_1,\cdots, z_n|y,x;T) = \frac{1}{4(n-1)!}\left(\frac{T}{4}\right)^{\frac{n-1}{2}} \e^{-\frac{\Delta^2}{4T}} \frac{d^{n-1}}{d\Delta^{n-1}} \Big\{ \e^{\frac{\Delta^2}{4T}} \erfc\left( \frac{\Delta}{2\sqrt{T}}\right)  \Big\}\,. \label{Rodrigues}
\end{equation}
Whilst it is simple to verify this reproduces \eqref{eqH23D1}, to convince oneself that it coincides with \eqref{eqH0Erfc} for arbitrary order the result of applying the derivatives can be written more explicitly in terms of the Hermite polynomials, denoted by $H_n$, as
\begin{align}
     {\H}_0^{(1)} (z_1,\cdots, z_n|y,x;T) = \frac{i^{n-1}}{4(n-1)!}\left(\frac{T}{4}\right)^{\frac{n-1}{2}}\Big\{& \erfc\left( \frac{\Delta}{2\sqrt{T}}\right)  H_{n-1}\left( \frac{-i\Delta}{2\sqrt{T}}\right)  \nonumber\\
     &+ \frac{2}{\sqrt{\pi}} \e^{-\frac{\Delta^2}{4T}} \sum_{k=1}^{n-1} \nCr{n-1}{k} i^{k}H_{k-1}\left( \frac{\Delta}{2\sqrt{T}}\right) H_{n-k-1}\left( \frac{-i\Delta}{2\sqrt{T}}\right)  \Big\}\,,
\end{align}
which define the polynomial coefficients of the complementary error function and Gaussian function mentioned above. If we now take the second derivative of the above representation the recurrence relation for Hermite polynomials ensures that it satisfies the differential equation (for brevity we indicate the order as an argument)
\begin{equation}
    \partial_{\Delta}^{2}  {\H}_0^{(1)}(n) + 8T\Delta \partial_{\Delta}  {\H}_0^{(1)}(n) -8T n {\H}_0^{(1)}(n) = 0\, .
    \label{eqHDiff}
\end{equation}
The generating formula (\ref{Rodrigues}) is a Rodrigues formula for the solutions of (\ref{eqHDiff}). Finally, to show that this coincides with the recursion relation in (\ref{eqH0Erfc}) we note that it is easy to prove by induction the following recursion relation for the iterated integrals of the complementary error function 
\begin{equation}
    2n\ierf{n}(z) + 2z\ierf{n-1}(z) - \ierf{n-2}(z) = 0\,, 
\end{equation}
which readily implies the differential equation (\ref{eqHDiff}).

With these explicit results it is now also a trivial exercise to confirm that the more general form of the operations that raise and lower the order, \eqref{convol}, \eqref{convolinv}, \eqref{Hstepn} and \eqref{Hstepninv} are satisfied by the $n$-hit function in one dimension (which is important because we constructed these functions using alternative relations). For $n \in \{1,2,3\}$ this can be verified on a case by case basis; the result for arbitrary order is again most easily proved by induction. This concludes our determination of the one dimensional generalised hit functions and, by extension, the $n$-hit function for arbitrary $n$ in any odd dimension. For the applications we have in mind to the Casimir effect, this is all we require. 

For the case of even dimensions, the recurrence relations and dimensional shift formulae imply that it suffices to obtain the $n=1$-hit function in $D=2$, say. To achieve this we invoke the integral representation \eqref{eqHintTime} which for $n=1$ and $D=2$ becomes 
\begin{align}
   \mathcal{H}_0^{(2)}(z |y,x;T) &= \frac{1}{(4\pi)^2}\int_0^T d\tau\, \frac{\e^{-\frac{\Delta_1^2 }{4\tau} - \frac{\Delta_2^2 }{4(T-\tau)}}}{\tau (T-\tau)} \,\theta(T) \nonumber \\
   &= \frac{1}{8\pi^2 T} K_0\left(\frac{\Delta_1 \Delta_2}{2T}  \right) \e^{\frac{\Delta^2}{4T}} \,\theta(T)\,, \label{Hit2D}
\end{align}
where we used \eqref{int4}. Thus we may, in principle, produce the infinite towers of hit functions of arbitrary order in even dimension by pure calculus.

\section{  Numerical Benchmarking }
\label{Benchmark}

To investigate the feasibility of the calculation of the propagator put forward by (\ref{Kpade}) we have chosen to test the proposal with specific examples in two different geometries, namely the $D=3$ spatial region $G$ with boundary $\partial G$, given either by a plane $\mathbb{R}^2$ or by a sphere $S^2$, where Dirichlet boundary conditions are imposed on the boundaries. For simplicity we fix the center of the sphere at the origin and the plane to be the $x$-$y$ plane passing through the origin. In the case of the sphere the region $G$ considered is the interior of the unit sphere while in the case of the plane the region $G$ consists of the half space in $\mathbb{R}^3$ bounded by the plane. 


In order to carry out a meaningful test we must compare the predictions given by (\ref{Kpade}) against the exact analytical expressions for the propagators (found in appendix \ref{exactprop}). This requires us to evaluate the $c_n$ up to a given order to determine a truncated approximation for the quotient of determinants. In practice we have calculated up to $n=3$, which corresponds to truncating the perturbative expansion at order $\lambda^4$. Evaluating $c_n$ as given by (\ref{cPade}) requires calculating multiple integrals over the boundary geometries that are in general difficult to determine analytically (except for some special cases, c.f. (\ref{exact1}), (\ref{Kplaneseries})),  especially for $n>1$. As such we resorted to numerical (Riemann) integration over the boundaries in such cases, although we should point out that despite this the calculation of (\ref{Kpade}) remains semi-analytic in that numerical techniques only enter the calculation to evaluate the coefficients $c_n$ defined in (\ref{cPade}). In our calculations this evaluation was done for $n=0,1,2,3$, using the Gauss-Kronrod quadrature method already built into {\it Mathematica 10.3.1} for the numerical integration over the surfaces $S$. We found that the relative error of this quadrature method is variable in our calculations but stays around $\epsilon_Q=0.1\%$; we discuss the error estimation in section \ref{Err}. The numerical evaluation of the integrals involved was greatly simplified by particular choices of coordinate systems to avoid spurious singularities which are detailed in appendix \ref{numericsApp}.

To obtain a graphical representation of the comparison we have plotted the dependence on the transition time, $T$, of the difference $K(x,y;T) - K_0(x,y;T)$ while keeping the points $x,y$ fixed in different cases. The difference $K(x,y;T) - K_0(x,y;T)$, the so called {\it scattering} or {\it vacuum-subtracted} propagator, is calculated in different manners for the purpose of comparison. First and foremost we compare all our results against the prediction given by the analytical (exact) expressions, that we have plotted in black. Secondly we plot the successive predictions, labelled $P_1^1, P_2^2,P_3^3$, for the vacuum-subtracted propagator given by formula (\ref{Kpade}) for $N=1,2,3$ for the truncated perturbative series. To be explicit, the quotients in these cases evaluate to

\begin{equation} \label{Padeexplicit}
P_1^1 (T)=-\frac{c_0^2}{c_1},\quad
P_2^2 (T)= \frac{c_1^3 + c_0^2c_3 - 2 c_0c_1c_2}{c_2^2 - c_1c_3},\quad 
P^3_3(T)= - \frac{c_2^4 - 3c_1c_2^2 c_3 +c_1^2c_3^2 + 2c_0 c_2 c_3^2}{c_3^3}\,,
\end{equation}
and will be plotted below as dashed green, orange and brown lines respectively. As will be seen the agreement we obtain is quite good already with just these three approximants but one might object that improving the accuracy further would require many more coefficients $c_n$ to be calculated. While this is indeed a viable way to proceed we can in fact improve the estimates without determining more coefficients by exploiting a simple observation: it turns out that in both geometries the sequence of $P_{N}^{N}$ alternates about the exact value (black solid line), behaviour that can be recognised even when the analytic result is unknown. We can use this to obtain a much better numerical estimate using convergence-acceleration methods.

As such we introduce the \textit{Shanks transformation} as a tool to accelerate convergence in the $\{P_N^N\}$ sequence. The Shanks transformation is particularly well suited to hasten convergence in alternating sequences \cite{bender2013advanced}, and it is this behaviour exhibited by $\{P_1^1,P_2^2,P_3^3\}$ that has prompted its application here. We call $S_1$ and $S_2$ the first and second iterated Shanks transformation of the sequence $\{P_N^N\}$ (see appendix \ref{PShApp}), explicitly given by
\begin{equation} \label{shanksPs}
    S_1 (T) =\frac{P_1^1 P_3^3 - (P_2^2)^2}{P_1^1 + P_3^3 - 2 P_2^2}, \quad 
    S_2 (T) =\frac{S_1 + P_1^1 - (P_2^2)^2}{S_1 + P_1^1 - 2 P_2^2}.
\end{equation}
In our plots we have displayed both $S_1$ and $S_2$ as solid blue and red lines respectively, from which we can appreciate a noticeable improvement. As expected, $S_2$ (in red) gives the closest fit to the analytical value (in black). Another favourable consequence of the alternating behaviour is that we can roughly estimate the systematic error using the Leibniz criterion discussed in Appendix \ref{Err}. We have included this estimation as error bars (in grey) centered around the $S_2$ curve. From a practical point of view we can consider $S_2$ to be the best estimate we obtained in both cases, using only the first four coefficients $c_n$ for its evaluation. 

The plots we obtained are given in figures \ref{SpherePlot1}, \ref{SpherePlot2} and \ref{PlanePlot}. They reveal that the accuracy achieved by this method is in most regions notably greater than our naive estimate would suggest, even if this estimate remains a tighter bound where the deviation is largest.

\subsection{Sphere}
Our first case of study is the interior region of a unit sphere (a unit ball) bounded by the surface $S= \partial G \simeq S^2$. Two cases were investigated where the vacuum-subtracted propagator was calculated with the same initial point $x$ but distinct final points $y$. We called these two cases {\it center-to-center} propagation and {\it center-off-center} propagation, corresponding to Cartesian coordinates $x=y=(0,0,0)$ and $x=(0,0,0),\, y= (0,0,r)$ respectively


\begin{figure}[h!]
\centering
  \includegraphics[width=0.65\textwidth]{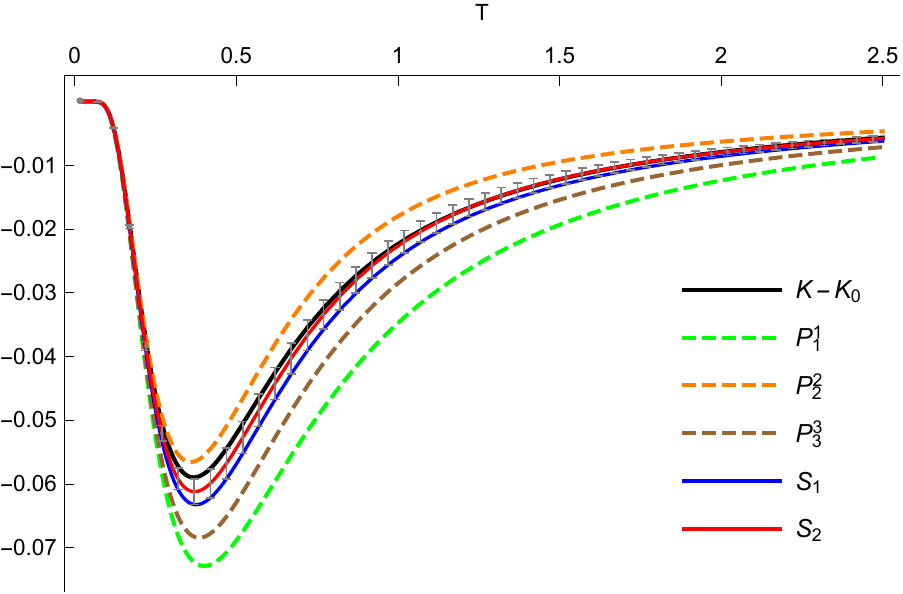}
  \caption{Center-to-center propagation amplitudes inside a sphere, comparing approximations of various orders against the known analytic results (black solid line)}
  \label{SpherePlot1}
\end{figure}

\begin{figure}[h!]
\centering
  \includegraphics[width=0.65\textwidth]{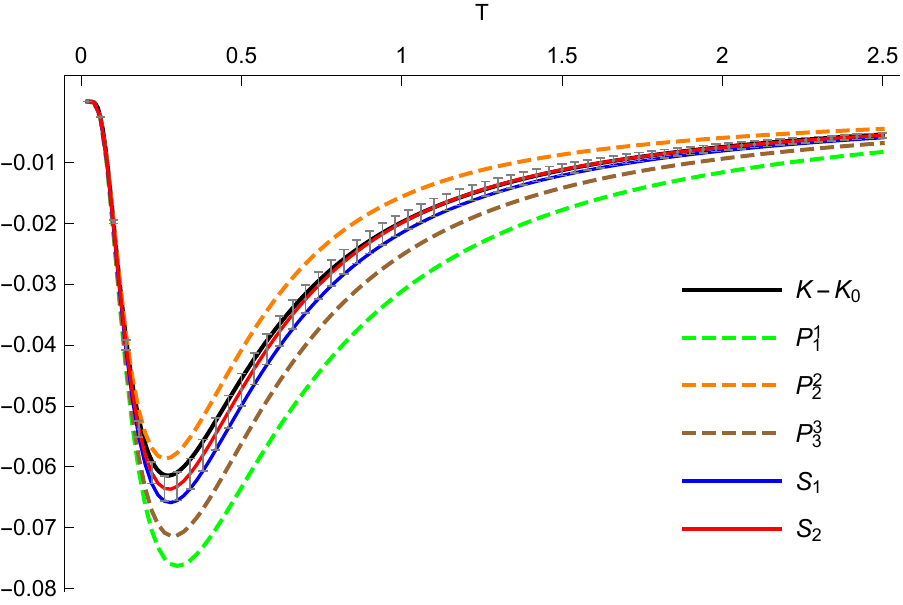}
  \caption{Center-off-center propagation amplitudes inside a sphere, showing agreement of the Pad\'e approximants to the analytic result (solid black line)}
  \label{SpherePlot2}
\end{figure}

We show the results obtained with numerical help for the multiple integration over the surface of the sphere in Fig. \ref{SpherePlot1} and Fig. \ref{SpherePlot2}. In both plots we have avoided the point $T=0$ (where the propagators are known to vanish), hence restricting to the interval $ 0.02 \leq T \leq 2.5$. For larger values of $T$ than those presented here we verified that the agreement improves asymptotically. 
In Fig. \ref{SpherePlot2} we have taken the parameter $r=0.7$, that is $y=(0,0,0.7)$ so that an appreciable deviation from the center-center propagation is observed.

For comparison to the exact result, the analytical value of the vacuum-subtracted propagator (in black) was calculated in both plots from the series representation developed in appendix \ref{exactprop} (see (\ref{KSPHERE})) by truncating it at $l_{max} = 3$ and $k_{max} = 8$. These values for truncation are optimal because, in fact, adding more terms slows down computation while producing no noticeable effect in our plots. This can be seen by examining $u_{4,1} \approx 8.18$, so that the largest ignored term is numerically $ \mathcal{O}(e^{-66.9 T})$ and the fast convergence rate of the series means the remaining terms do not contribute substantially. Other calculations are of course possible where $x$ and $y$ are completely arbitrary points (not necessarily colinear with the center of the sphere) and based on our numerical calculations we expect that a similar agreement would be found.

\subsection{Plane}
 
The same comparison was done for the region (half-space) bounded by an infinite plane, in this case $S =\partial G \simeq \mathbb{R}^2$; we investigated propagation from a point ${\bf p} = (0,0,d)$ to itself, called ${\bf p}$ to ${\bf p}$ propagation. An important difference with the previous section is that the plane is infinite, and hence in the numerical integration demanded by the calculation of the coefficients $c_n$ a finite region $G$ of the plane must be chosen that is large enough to introduce only small errors (this is possible thanks to the exponential fall off of the hit functions, (\ref{Hit3D}), with distance from the endpoints) -- this procedure is detailed in Appendix \ref{numericsApp}. 

\begin{figure}[h!]
\centering
  \includegraphics[width=0.6\textwidth]{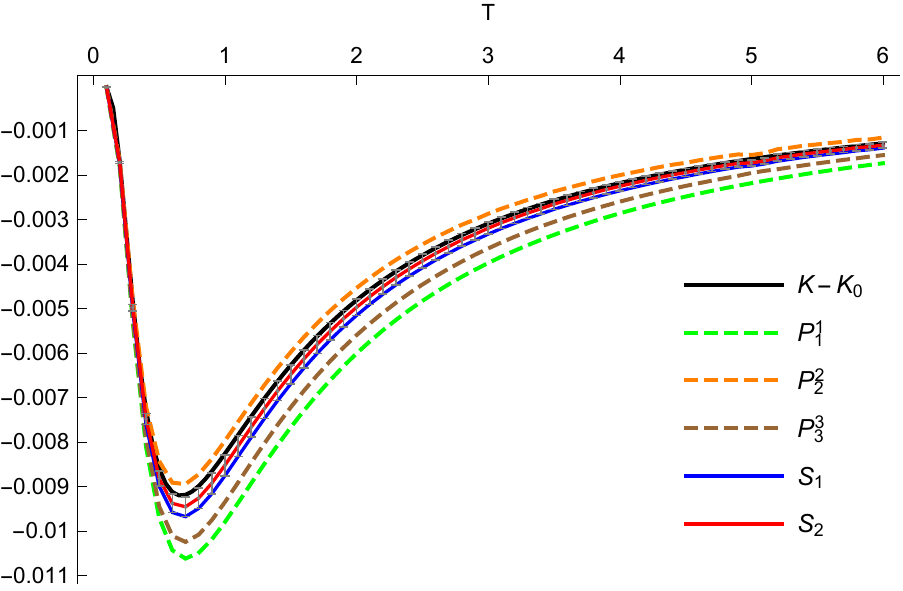}
  \caption{Comparing {\bf p} to {\bf p} propagation amplitudes in the half-space against the semi-analytic result (\ref{KSPHERE}).}
  \label{PlanePlot}
\end{figure}

Since the plane is infinite, in considering ${\bf p}$ to ${\bf p}$ propagation the only length scale is $d$, the distance from ${\bf p}$ to the plane, therefore no generality is lost if we assume the numerical value $d=1$ that we have chosen for simplicity in our numerical calculations. In Fig. \ref{PlanePlot} we chose for clarity the interval $0.1 \leq T \leq 6$ since, as in previous plots, the agreement improves for larger values of $T$.

One can see that Fig. \ref{PlanePlot} shows how the results obtained for the plane have a comparable degree of accuracy to those found in the sphere. We interpret this good agreement as indicative that if further levels of accuracy are calculated in (\ref{Kpade}) one can obtain a sequence of approximants that will converge to the desired difference of propagators. Naturally, since we have only considered three different cases presented in figures \ref{SpherePlot1},\ref{SpherePlot2} and \ref{PlanePlot} based on truncation of (\ref{Kpade}) we can only consider this as evidence to support our claimed conjecture.

\section{Conclusions}

In this article we have generalised the so-called ``hit function,'' introduced in previous work, to a more general integral transform of the quantum mechanical propagator: this $n$-hit function gives the relative contribution to the kernel involving propagation between its arguments through $n$ (the order) intermediate points and contains the full information of the of the quantum system in question, being indeed a kind of generalised Green function for the Schr\"{o}dinger operator as illustrated by (\ref{eqHGreen}). The $n$-hit function is notionally defined as a constrained path integral but we have developed various formulas, including (Riemann) integral representations, of this hit function that prescribe how to determine it from knowledge of the kernel.

We have also presented two ways in which quantum mechanical propagators can be constructed from the general hit function. The first, equation \eqref{RelNormalisation}, is the direct inverse transformation recovering the unconstrained kernel (that extends to the case that the hit function is computed for a non-trivial kernel, i.e. in the presence of a background potential). The second, outlined in \eqref{Khitseries} and developed in sections \ref{QuantP} and \ref{PadeRepn} , shows how the hit functions of different orders can be formed into a series representation of the propagator in localised potentials or Dirichlet boundary conditions on different geometries. For the case of Dirichlet boundary conditions, our method combines nicely with Pad\'e extrapolation allowing us to access the strong coupling regime of the series representation with remarkable success. While this has been illustrated for the case of propagation in the presence of a conducting plate and in the interior of a conducting spherical shell where we were able to show close agreement with analytical results obtained by conventional methods, the approach presented here can be used to calculate numerically the propagator/heat kernel in a bounded region with an irregular shape where an analytical expression is not viable. The proposal we make here could be considered as outlining a new form of perturbative expansion for application in the strong coupling regime -- as illustrated in (\ref{Kpade}) this suggestion estimates the kernel as a quotient of matrix determinants involving matrices of increasing size, using the Pad\'e method to extrapolate to large values of the coupling parameter so that in this sense this can be considered a non-perturbative method based on perturbation theory. We have also shown that the convergence to this region of parameter space can be sped up by taking advantage of the (iterated) Shanks transformation. In section \ref{Vpot} we point out that this proposal can easily be adapted to non-localised static potentials of more general shape. 

As part of the development of this new approach to calculating the kernel we have also presented some general properties of these hit functions and derived relationships between the functions in different dimensions and of different order. It is important to emphasise that these functional relations imply that knowledge of one hit function (for some order) in one even- and one odd-dimensional space is, in principle, sufficient to determine \textit{all} hit functions. To this end we have supplied the explicit form of all $n$-hit functions in $D=1$ (and in fact also for $D=3$) and calculated the $1$-hit function in $D=2$ from which all other hit functions can be determined. Of course, the difficult parts of the process may be in successfully applying the identities of section \ref{secIds} to obtain explicitly the hit functions of different orders in the required dimension and then compute the appropriate integrals for the geometry of the system in question.

Before finishing we consider future work and possible extensions of the results presented here. 
As discussed above, especially in the context of our applications to the planar and spherical geometries in section \ref{Benchmark}, the generalised hit function finds a natural application to studies of the Casimir energies in different geometries based on the first quantised representation of the effective action, which make for difficult calculations using standard techniques except in simple or special cases. The results of this application will be published in a later communication for some Casimir geometries of phenomenological interest to estimate the vacuum energy density and Casimir force. We estimate that the numerical precision we have attained is enough for such applications at a reasonable computational cost.

Longer term, other extensions include the incorporation of internal angular momentum -- where the hit function will distinguish the contribution to the kernel from propagating modes for different spin degrees of freedom -- and the generalisation to relativistic particles and the Minkowski space path integral which connects to quantum field theory via the Worldline Formalism. 
The hit function approach can also be used in scattering problems where the effect of a localised potential on an incoming particle is to be accounted for, so we believe there is a possibility of using a relativistic generalisation of the hit function to treat high-energy scattering problems in QFT in a new manner. We would also like to point out that scattering of electromagnetic waves can be investigated using the presented method. The main difficulty here will be to implement the correct boundary conditions on the vector potential $A(x,t)$ and the transversality of the electromagnetic fields. The Klein-Gordon propagator can already be obtained from the non-relativistic Schr\"odinger propagator by means of the worldline formalism through an integral transform \cite{ChrisRev, UsRep}. Since the approach presented here provides a pathway to non-perturbative calculations it would be of interest to find out the scope of validity of this technique. 

Yet another application can be found if the hit function is interpreted within the framework of the heat/diffusion equation. Here the results of the present paper allow one to solve the heat equation or the diffusion equation in bounded regions numerically, a problem of present day relevance in the study of Brownian motion. We also hope that the new method we have introduced might stimulate interest within the framework of heat-kernel techniques where it might be of some relevance in the calculation of Seeley-DeWitt coefficients \cite{vassilevich2003heat}. 

In light of the success we have obtained in calculating point-to-point propagation amplitudes we might use this method to calculate the density of states $\rho (E) = \frac{1}{\pi} \Im \int d^D x\, \hat{K}(x,x;-iE)$ for a quantum particle trapped inside a cavity and investigate the effects of changes in the shape of the cavity. For a many particle quantum system, for instance a quantum gas at finite temperature $1/\beta$, a multi-particle generalisation would require the use of the coordinate multi-vector $\{x\} = \{x_1,x_2,\cdots, x_\mathcal{N}\}$; the method outlined here then would estimate the multi-particle point-to-point propagator $K (\{x\},\{x\};T)$ so that the partition function of the system could be calculated as $Z(\beta) = \int d^D\{x\}\, K (\{x\},\{x\};\beta)$. Its practical use will depend on whether the required integrations can be carried out.

Finally, we allow ourselves to make a more speculative observation regarding Feynman's original construction of the path integral. The classic construction, as expounded in \cite{feynman2010quantum}, for example, uses the intuition supplied by the Young's slits experiment to motivate dividing propagation into infinitesimal steps, integrating the intermediate points along the trajectory over all space. One could interpret the large $n$ limit of the generalised hit function we define here as a concrete realisation of this splitting procedure, following which the continuum path integral could be defined as the $n \rightarrow \infty$ limit of (\ref{RelNormalisation}). Of course the precise nature in which this limit is determined requires further analysis to make this interpretation more precise but as a first attempt one could consider reproducing the interference pattern for double slits of finite size using the $n=1$ hit function integrated over a finite volume. It also follows from this analysis that the $n$-hit function will be useful in investigating the quasi-classical evolution of a quantum system when the classical trajectory goes through the prescribed space-time points $(z_1,\tau_1),\cdots,(z_n,\tau_n)$.

\section*{Acknowledgements}
JPE and IH are pleased to thank CONACyT for financial support via a ``Ciencia de Fronteras'' project \#\!\! 170724. JPE acknowledges additional funding through a Universidad Michoacana CIC project. IH thanks Marco Sch\"afer for insightful exchanges. The authors thank Christian Schubert and Axel Weber for helpful discussions and Fiorenzo Bastianelli and Petr Jizba for comments on the manuscript and various suggestions. 

\appendix
\section{3 dimensional $n$-hit function}
\label{apH3D}
We outline the calculation of the closed form for the $n$-hit function in $D=3$. In what follows $\alpha, \beta \in \mathbb{C}$ and $\Re(\alpha) \geq 0, \Re(\beta) \geq 0$. Consider first the following integral

\begin{equation} \label{int3}
\int_{0}^{\infty} \exp\left( -\alpha x^2 - \frac{\beta}{x^2} \right)dx = \half \sqrt{\frac{\pi}{\alpha}}e^{-2 \sqrt{\alpha}\sqrt{\beta}}\, .
\end{equation}
This identity can be proved by changing to the variable $y = \sqrt{\alpha} x - \frac{\sqrt{\beta}}{x}$ which renders (\ref{int3}) in Gaussian form after using contour integration. Alternatively one may use a complex Cauchy-Schl\"omilch transformation in the integrand by changing variable to $\xi = \frac{1}{x} \frac{\sqrt{\beta}}{\sqrt{\alpha}}$ and then use Cauchy's theorem to deform the contour so that the $\xi$ integration is shifted to run on the real axis; finally replacing $\xi \to x$ and averaging both integrals casts the original integral into a Gaussian integral leading to the RHS of (\ref{int3}). This result extends the integral {\it 3.325} in \cite{gradshteyn2007table} to complex values of $\alpha$ and $\beta$. We have also found it useful to compute the following integrals  

\begin{eqnarray} \label{int1}
\int_0^T \exp\left( -\frac{\alpha}{T-\tau} - \frac{\beta}{\tau} \right) \frac{d\tau}{\sqrt{\tau} (T-\tau)^{\frac{3}{2}}} &=& \sqrt{\frac{\pi}{\alpha T}} e^{-\frac{(\sqrt{\alpha} + \sqrt{\beta})^2}{T}} \,, \\ \label{int2}
\int_0^T \exp\left( -\frac{\alpha}{T-\tau} - \frac{\beta}{\tau} \right) \frac{d\tau}{ (\tau(T-\tau))^{\frac{3}{2}}} &=& \sqrt{\frac{\pi}{T^3}} \frac{\sqrt{\alpha} + \sqrt{\beta}}{\sqrt{\alpha}\sqrt{\beta}}e^{-\frac{(\sqrt{\alpha} + \sqrt{\beta})^2}{T}} \,,\\
\label{int4}
\int_0^T \exp\left( -\frac{\alpha}{T-\tau} - \frac{\beta}{\tau} \right) \frac{d\tau}{\tau (T-\tau)}
&=& \frac{2}{T} K_0 (2\sqrt{\alpha \beta}/T)e^{-(\alpha + \beta)/T}\,.
\end{eqnarray}
  To prove (\ref{int1}) we change variable to $x= \sqrt{\frac{\tau}{T-\tau}}$ which reduces the problem to (\ref{int1}) to that of calculating (\ref{int3}). Finally, by acting with $-\frac{\partial }{\partial \beta}$ we can relate (\ref{int1}) to (\ref{int2}). Particular cases of (\ref{int3}),(\ref{int1}) and (\ref{int2}) are found in the appendix of \cite{feynman2010quantum} when $\alpha$ and $\beta$ are purely imaginary. One may now solve the integrals in representation (\ref{HitTime}) by substituting (\ref{K0inD}) with $D=3$ and repeated use of (\ref{int2}). The key point is that $D=3$ is special, since it is the only dimension where the $\tau$ integral of the product of two free propagators is again proportional to a free propagator, this allows one to obtain directly
\begin{equation}
\H_0 (z_1 ,z_2, \cdots, z_n| y,x ; T)=\frac{1}{(4\pi)^{\frac{2n+3}{2}}} \frac{e^{- \frac{\Delta^2}{4T} }}{T^{3/2}} \prod_{j=1}^{n}
\left( \Delta_{j+1}^{-1} + \left(\sum_{k=1}^{j} \Delta_k \right)^{-1} \right)\,,
\end{equation}
that can be simplified down to (\ref{Hit3D}).

\section{Constrained path integral}
\label{appPI}

In this appendix we provide two approaches to prove identity (\ref{deltas}). One is based on a path decomposition induced by a natural proper time interval partition and the other based on the Fourier representation of Dirac $\delta$-distributions and generating function techniques.

\subsection{First Approach}

Given boundary conditions $x(0) = x, x(T)=y$ a path integration with an action functional $S(0,T) = \int_{0}^T \mathcal{L} \,dt$ over all trajectories that satisfy these boundary conditions can always be split for any value $s$ such that $0<s<T$ as

\begin{equation}
    \int_{x(0)=x}^{x(T)=y} \mathscr{D} x \,  \e^{-S[0,T]} = 
\int d^D w  \int_{x(0)=x}^{x(s)=w} \mathscr{D} x \,  \e^{-S(0,s)}  \int_{x(s)=w}^{x(T)=y} \mathscr{D} x \,  \e^{-S(s,T)}\,.
\end{equation}
Consider now the identity found in \cite{PvHz,feynman2010quantum} for $0<\tau<T$

\begin{equation} \label{pathsplit}
    \int_{x(0)=x}^{x(T)=y} \mathscr{D} x  \, \e^{- S_0 (0,T) }\,\delta^D (x(\tau) - z) =
    K_0 (y,z;T-\tau ) K_0 (z,x; \tau)\,,
\end{equation}
where $S_0 (0,T) := \int_0^T dt  \frac{\dot{x}^2}{4}$. Given now a set of ordered proper times $0\leq \tau_1\leq\tau_2\leq \cdots \leq \tau_n \leq T$ we can always find a set $\mathcal{S}=\{ s_k : 0 \leq k \leq n \}$ so that  $0:=s_0 \leq \tau_1\leq s_1 \leq  \tau_2 \leq s_2 \leq \cdots \leq s_{n-1}\leq \tau_n \leq s_{n}:=T$
and evaluate $\mathcal{I}_{n}$ by splitting the $[0,T]$ interval precisely at each $s_k$, that is

\begin{equation}
  \mathcal{I}_{n} = \prod_{k=1}^{n} \int d^D w_k  \int_{x(s_{k-1})=w_{k-1}}^{x(s_k)=w_k}\!\!\!\!\!\! \mathscr{D} x \, \e^{- S_0[s_{k-1}, s_k]}\, \delta^D (x(\tau_k) - z_k)\,,
\end{equation}
where we defined $w_0 := x, w_{n} := y$. Now use (\ref{pathsplit}) repeatedly to solve for each term to arrive at

\begin{equation}
    \mathcal{I}_{n} = \prod_{k=1}^{n} \int d^D w_k \, K_0 (w_k,z_k ; s_k -\tau_k)K_0 (z_k,w_{k-1};\tau_{k}-s_{k-1})\,.
\end{equation}
The integrations can now be performed directly using the reproducing kernel property; the result is independent of $\mathcal{S}$ and is in fact (\ref{deltas}).

\subsection{Second Approach}

Here we give a brief proof of the result (\ref{eqIn}) by explicit evaluation of the path integral. Using the Fourier decomposition of the $\delta$-functions under the integral we can write
\begin{equation}
    \mathcal{I}_{n} = \prod_{i=1}^{n}\int \frac{d^{D}k_{i}}{(2\pi)^{D}} \e^{i\sum_{i = 1}^{n}k_{i}\cdot z_{i}} \int_{x(0) = x}^{x(T) = y}\hspace{-1.5em} \mathscr{D}x(\tau) \, \e^{-\int_{0}^{T} d\tau \big[ \frac{\dot{x}^{2}}{4} + i j(\tau) \cdot x(\tau) \big]}\,, 
\end{equation}
where $j(\tau) = \sum_{i=1}^{n} k_{i}\delta(\tau - \tau_{i})$. Completing the square in the path integral exponent and evaluation of the Gaussian integral lead to
\begin{equation}
    \mathcal{I}_{n} = (4\pi T)^{-\frac{D}{2}}\e^{-\frac{(x-y)^{2}}{4T} } \prod_{i=1}^{n}\int \frac{d^{D}k_{i}}{(2\pi)^{D}} \e^{\sum_{i, j =1}^{n} k_{i} \cdot k_{j} \Delta_{ij} + i \sum_{i=1}^{n} k_{i}\cdot (z_{i} - \hat{x}_{i} )}\,,
\end{equation}
where $\Delta_{ij} := \Delta(\tau_{i}, \tau_{j}) = \frac{1}{2}\I \tau_{i} - \tau_{j}\I - \frac{1}{2}(\tau_{i} + \tau_{j}) + \frac{\tau_{i}\tau_{j}}{T}$ is a ``worldline Green function'' for the free particle subject to Dirichlet boundary conditions on $[0, T]$ and $\hat{x}_{i} := \hat{x}(\tau_{i}) = x + (y-x)\frac{\tau_{i}}{T}$ is a parameterisation of the straight line path between $x$ and $y$ in time $T$. 

The (Riemann) integrals over the $k_{i}$ are also Gaussian and evaluate to give
\begin{equation}
    \mathcal{I}_{n} = (4\pi T)^{-\frac{D}{2}}\e^{-\frac{(x-y)^{2}}{4T} } (4\pi)^{-\frac{nD}{2}} \textrm{det}^{-\frac{D}{2}}(-\Delta)\e^{-\frac{1}{4} \sum_{i,j=1}^{n}(z_{i}-\hat{x}_{i})\Delta^{-1}_{ij} (z_{j} - \hat{x}_{j})}\, .
    \label{eqInDelta}
\end{equation}
To confirm that this coincides with the product of propagators in (\ref{eqIn}) we fix the ascending order used throughout the main text. Then
\begin{align}
    \textrm{det}(-\Delta) &= \frac{1}{T}\prod_{i = 1}^{n+1} (\tau_{i} - \tau_{i-1})\, , \\
    \sum_{i,j=1}^{n}(z_{i}-\hat{x}_{i})\Delta^{-1}_{ij} (z_{j} - \hat{x}_{j})&= \frac{(z_{n} - \hat{x}_{n})^{2}}{T-\tau_{n}} + \sum_{i = 1}^{n-1} \frac{(z_{i} - z_{i+1} - \hat{x}_{i+1} + \hat{x}_{i})^{2}} {\tau_{i} - \tau_{i+1}} + \frac{(z_{1} - \hat{x}_{1})^{2}}{\tau_{n}} \nonumber \\
    &=  \frac{(z_{n} - y)^{2}}{T - \tau_{n}} + \sum_{i, j = 1}^{n-1} \frac{(z_{i} - z_{i+1})^{2}}{\tau_{i} - \tau_{i+1}} + \frac{(z_{1} - x)^{2}}{\tau_{1}} - \frac{(x-y)^{2}}{T}\, .
\end{align}
Substituting these results into (\ref{eqInDelta}) we recover the product in (\ref{deltas}) for the chosen ordering. For evaluation of a similar path integral (essentially the Fourier transform of the one presented here) see \cite{Bastianelli:2014bfa, Fujiwara:1981rf, Polyakov:1987ez}.

\section{Exact propagators} \label{exactprop}
Here we show how the exact propagators were derived analytically for the interior region bounded by a spherical $2$-dimensional shell in a $3$-dimensional Euclidean space and for the half $3$-dimensional Euclidean space bounded by a $2$-plane; these propagators were obtained following the standard methods of mode summation/spectral decomposition and the image method respectively.

\subsection{Propagator inside the sphere}

The propagator in question inside the unit sphere satisfies the Green equation
\begin{equation} \label{GreenK}
    \left( \frac{\partial}{\partial t} - \nabla^2 \right) K (x,x';t) = \delta (t) \delta^{3} (x-x')\,,
\end{equation}
along with the Dirichlet boundary conditions
\begin{equation} \label{boundB}
K(x,x'; 0) = \delta (x-x'), \qquad  K(x,x'; t) = 0,\mbox{ for } t<0,\qquad K(x,x'; t) = 0 \mbox{ for } |x| =1. 
\end{equation}
We solve (\ref{GreenK}) using separation of variables, which naturally leads to the following structure for the solution
\begin{equation} \label{Ksep}
    K(x,x'; t) = \sum_{\mu, \ell, m} A_{\ell,m,\mu}(x') e^{-\mu t} j_{\ell}(\sqrt{\mu} r)Y_{\ell m} (\Omega) \theta (t)\,,
\end{equation}
where $\Omega = (\theta,\phi)$ are the spherical angle coordinates of $x$, $r=|x|$, $j_{\ell}$ are the spherical Bessel functions, $Y_{\ell m}$ are the spherical harmonics, and the coefficients $A_{\ell,m,\mu}(x')$ are fixed by the boundary conditions (\ref{boundB}). First we notice that necessarily $\mu = u_{\ell,k}^2$ being $u_{\ell,k}$ the zeroes of $j_{\ell}(r)$ for $k=1,2,\cdots$, therefore the values of $\mu$ are labeled by the index pair $(\ell,k)$ A way to proceed now is to observe that the orthogonality relation
\begin{equation}
    \int_0^1 x^2 j_{\ell} (x u_{\ell,k})j_{\ell} (x u_{\ell,k'}) dx = \frac{\delta_{kk'}}{2} (j_{\ell+1}(u_{\ell,k}))^2
\end{equation}
implies in turn the completeness relation
\begin{equation}
    \frac{\delta (r-r')}{r^2} = 2 \sum_{k=1}^{\infty} \frac{j_{\ell} (r u_{\ell, k}) j_{\ell} (r' u_{\ell,k})}{j^2_{\ell+1} (u_{\ell,k})}\,,
\end{equation}
which taken together with the completeness relation for the spherical harmonics (where $\cos \gamma = \hat{x} \cdot \hat{x}'$),
\begin{equation}
\sum_{\ell=0}^{\infty}\sum_{m= -\ell}^{\ell} Y_{\ell m}^* (\Omega') Y_{\ell m}(\Omega) = \sum_{\ell =0}^{\infty} \frac{2\ell +1}{4\pi} P_{\ell} (\cos \gamma)= \delta (\Omega - \Omega')\,,
\end{equation}
and setting $t=0$ in (\ref{Ksep}) yield $A_{\ell,m,\mu}(x') = Y^*_{\ell m} (\Omega') j_{\ell}(r' u_{\ell,k})/j_{\ell+1}^2 (u_{\ell,k})$, finally

\begin{equation} \label{KSPHERE}
   K(x,x';t) = \sum_{\ell =0}^{\infty} \sum_{k=1}^{\infty}\frac{2\ell +1}{2\pi} \frac{ j_{\ell}(r u_{\ell,k})j_{\ell}(r' u_{\ell,k})}{j_{\ell+1}^2 (u_{\ell,k})}
   P_{\ell}(\cos \gamma)e^{-u_{\ell,k}^2 t}\,\, \theta (t) \,,
\end{equation}
which can be cut off at appropriate values of $l$ and $k$.
\subsection{Propagator in half space}

For  the  simple geometry of Dirichlet boundary conditions along an infinite plane the analytic calculation of the kernel is trivialised by the method of images. Indeed, denoting by $\R{{\bf p}}$ the point corresponding to the reflection of ${\bf p}$ in the plane, we may write the kernel on the half-space ($y$ and $x$ are on the same side of the plane, as the image method requires, for $y$ and $x$ on different sides the Dirichlet boundary conditions imply $K(y,x  ; T) =0$)
\begin{equation} \label{KPLANE}
    K(y,x  ; T) =  K_0 (y, x; T)  - K_0 (\R{y}, x; T)\,,
\end{equation}
which, by construction, vanishes for points $y = \R{y}$ (or, by symmetry of the kernel $x = \R{x}$)  on the surface of the plane and satisfies the appropriate Green equation on the half-space. In a path integral calculation, this result follows from  noting that the contribution from trajectories defined in $\mathbb{R}^{D}$ that enter the forbidden region as they propagate to $y$ are cancelled by those that travel to the image point $\R{y}$, leaving only those that remain in the correct half-space (including various reflections off the plane) -- see \cite{PIHalfSpace} for more information. Our calculation with the hit function will implement this cancellation by explicitly excluding paths that penetrate the boundary.

 \label{AppD}

\section{Sphere and plane numerics} \label{numericsApp}

The main difficulty in the numerical evaluation of the coefficients $c_n$ in (\ref{cPade}) for the three-dimensional case is the presence of the denominator $1/\prod_i \Delta_i $ in the integrand (\ref{cPade}), since as this is integrated numerically it produces spurious divergences when adjacent points $z_i$ and $z_{i+1}$ become coincident. These divergences are only artificial and they are easily removed by using an appropriate coordinate system. This is in general the case; on a sufficiently regular surface $S$ a coordinate system can always be defined so that all such spurious divergences disappear.  We show how this is attained in our cases of study and give more details on the numerics that led to the results reported in section \ref{Benchmark}.

\subsection{Sphere}

To eliminate the spurious divergences in $c_{n-1}$ we introduce a coordinate system to evaluate (\ref{cPade}), namely
\begin{equation} \label{GINT}
  (-1)^{n} \int_{\partial G} d\Omega_1 \cdots \int_{\partial G} d\Omega_{n}   \,  \H_0 (z_1,\cdots,z_{n}|y,x;T)\,,
\end{equation}
where $d\Omega_i$ is the solid angle measure for the sphere $\partial G$ parametrised by coordinates for $z_i$. These coordinates are specified as follows. First let the last point $z_n$ have the standard $(\theta_n, \phi_n)$ spherical coordinates with origin at the center of the sphere, choosing the orientation for the $z$ (azimuthal) axis so that the point $y= (0,0,r)$ lies along it. Next consider the point $z_{i}$ for $1 \leq i \leq n-1$ and assign it the spherical coordinates $(\theta_{i}, \phi_{i})$. Again the origin of this $i$-th coordinate system coincides with the center of the sphere but the azimuthal axis now goes through $z_{i+1}$ so that the polar angle $\theta_{i}$ is actually measured with respect to the radius that passes through $z_i$. 

\begin{figure}[h!]
\centering
  \includegraphics[width=0.35\textwidth]{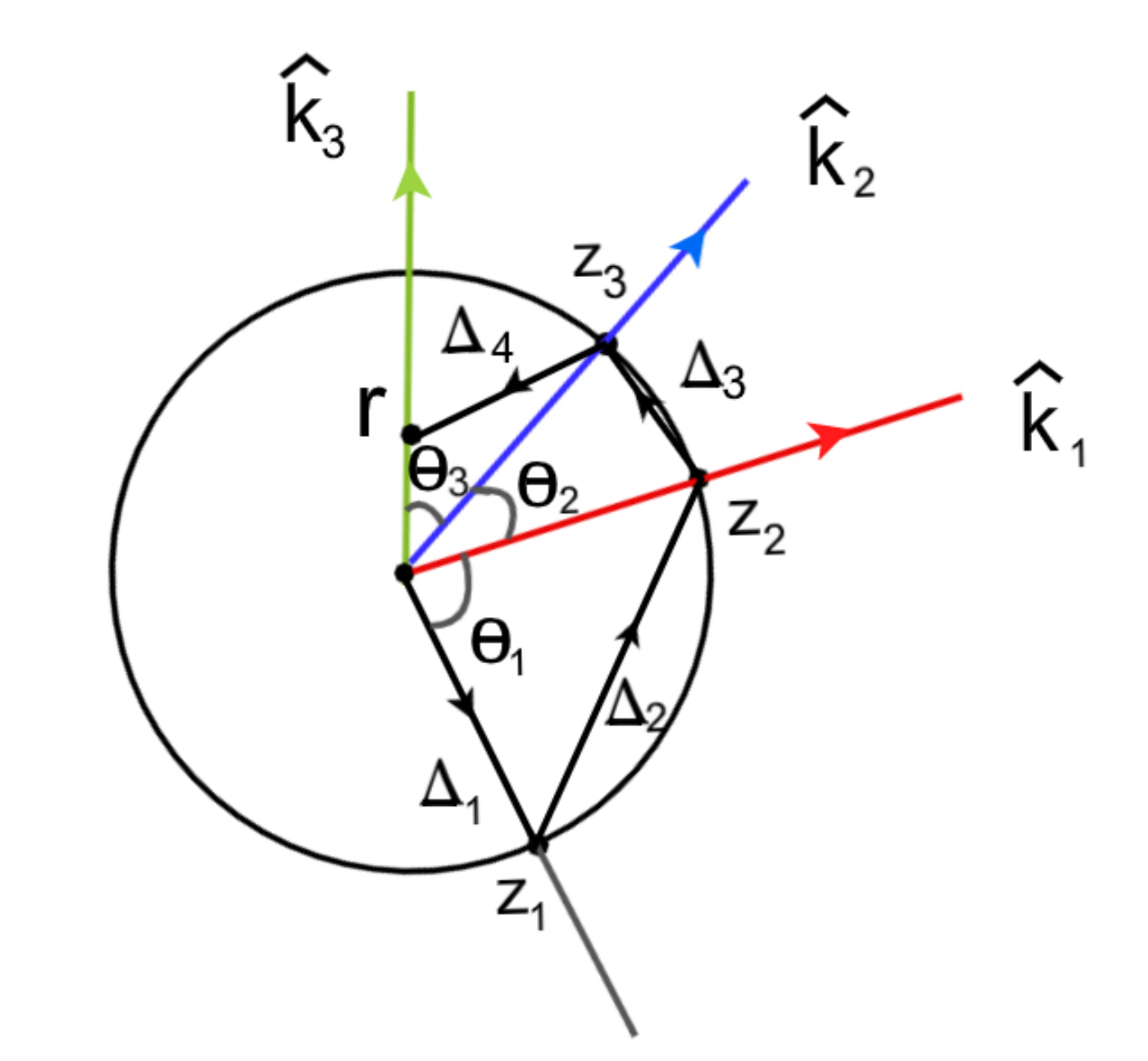}
  \caption{Illustration of the coordinate system on the sphere, where each new vertex $z_i$ is described so that the azimuthal direction $\hat{k}_i$ goes through $z_{i+1}$. The polygonal path starts at the center of the sphere and ends at a point located at the distance $r$ from the origin.}
  \label{CoordDiag}
\end{figure}

This has been illustrated in Fig. \ref{CoordDiag} for $n=3$, where the first choice is made for $z_3$ whose azimuthal axis is $\hat{k}_3$ (in red). Next the axis is rotated to the direction $\hat{k}_2$ (in blue) and finally to $\hat{k}_1$ (in green). The polar coordinates of the point $z_i$ are described with respect to the polar system whose azimuthal axis is $\hat{k}_i$. This peculiar choice of coordinates allows one to perform all the $\phi_i$ integrations straightforwardly so that only the polar integrations over $\theta_i$ remain to be done. These are greatly simplified through the change of variables

\begin{equation}
\xi_i = \sin(\theta_i/2), \quad 1 \leq i \leq n-1, \quad 
\xi_n = \cos \theta_n\,.
\end{equation}
In particular the lengths of segments and the total length of the polygonal path become
\begin{equation}
\Delta_1 = 1,\quad \Delta_{n+1}=\sqrt{1+r^2-2r \xi_n},\quad \Delta_i = 2\xi_i, \quad 2 \leq i \leq n, \quad \Delta = 1 +\sqrt{1+r^2-2r\xi_n} + 2\sum_{i=1}^{n-1} \xi_i\, .
\end{equation}

In effect we can apply the following substitution rules when calculating (\ref{GINT})
\begin{eqnarray}
\int \frac{d\Omega_i}{\Delta_{i+1}} &\to& 4\pi \int_0^1 d\xi_i ,\quad \,,
1\leq i \leq n-1 \nonumber \\
\int d\Omega_n &\to& 2\pi \int_{-1}^{1} d\xi_n  \label{centertor}\,,
\end{eqnarray}

These rules simplify greatly the calculation, for example in the center-center propagation ($r=0$) the $n$-th coefficient becomes
\begin{equation}
    (-\lambda)^n \frac{2}{(4\pi T)^{3/2}} \int_0^1 d\xi_1 \cdots \int_{0}^1 d\xi_{n-1} 
    \left( 1 + \sum_{i=1}^{n-1}\xi_i \right) e^{-\left(1+ \sum_{i=1}^{n-1}\xi_i \right)^2 /T}\,,
\end{equation}
and the whole series can be calculated analytically as

\begin{equation} \label{exact1}
    K(0,0;T) = \frac{1}{(4\pi T)^{3/2}}- \lambda \frac{2}{(4\pi T)^{3/2}}e^{-1/T}  + \lambda^2 \frac{1}{(4\pi T)^{3/2}} T \left( e^{-1/T}-e^{-4/T}\right) + \cdots 
\end{equation}
Note that in the general case of center-off-center propagation the original $2n$ integrations are reduced to $n$.

\subsection{Plane}

In calculating the terms of the perturbative series (\ref{Khitseries}) for the half space bounded by a plane the first two terms can be determined analytically in the case of propagation from a point $\p$ to itself after proper time $T$ has elapsed 
\begin{equation} \label{Kplaneseries}
K(\p,\p;T) =  K_0 (\p,\p;T) - \lambda \frac{1}{16\pi T} \erfc( d/\sqrt{T}) + \lambda^2 \cdots
\end{equation}
where $d$ is the distance from $\p$ to the plane. There is no loss of generality in taking $\p = (0,0,d)$ and the plane to be the $x$-$y$ plane passing through the origin, numerically we take $d=1$. 

Higher order terms require numerical integration to be evaluated, so to that end let us introduce polar coordinates when integrating over the plane  with an origin coincident with the origin of the original coordinate system. As a first example consider the $\lambda^2$ term in (\ref{Kplaneseries}), take $z_1 =(\rho_1,\phi_1)$ so that the angle $\phi_1$ is measured with respect to the x axis on the plane, then take $z_2 = (\rho_2,\phi_2)$ where now $\rho_2$ is the distance measured with respect to $z_1$, and  the angle $\phi_2$ is measured with respect to radius $\rho_1$ so that 
\begin{equation}
\Delta_1 =\sqrt{d^2 + \rho_1^2},\quad
\Delta_2 = \rho_2,\quad
\Delta_3 = \sqrt{\rho_1^2 + \rho_2^2 -2\rho_1\rho_2 \cos \phi_2 + d^2}\,,
\end{equation}
and the measures on the plane become $d\sigma_1 = \rho_1 d\rho_1 d\phi_1, \, d\sigma_2 = \rho_2 d\rho_2 d\phi_2$ this choice of coordinates removes the spurious singularity in the integrand $1/\Delta_2$ as $\rho_2 \to 0$ simplifying the convergence of numerical integration. The coordinate $\phi_1$ can be integrated over directly resulting in a $2\pi$ factor and a well behaved integrand. We found it convenient to introduce a cut-off in the radial distance being integrated so that in practice $\rho_1,\rho_2 \in [0, \rho_{max}]$, with $\rho_{max} = 20$ in calculating $c_1$ and $\rho_{max} = 10$ in calculating $c_2$ and $c_3$. Increasing the integration cut-off $\rho_{max}$ produces numerically identical values within the quadrature accuracy. 
\begin{equation}
    c_1 = \frac{(2\pi)}{(4\pi)^{7/2}T^{3/2}}\int_0^{\rho_{max}}\!\!\!\!\!\!\!\! d\rho_1 
    \int_0^{2\pi}\!\!\!d\phi_2 \int_0^{\rho_{max}} \!\!\!\!\!\!\!\!d\rho_2
    \,\frac{ \,\e^{-\Delta^2/4T}\rho_1 \Delta}{\sqrt{(d^2+\rho_1^2)(d^2 + \rho_1^2 + \rho_2^2- 2\rho_1\rho_2 \cos \phi_2)}}\,.
\end{equation}

For $n>2$ we must introduce additional points $z_k$. Now the procedure is changed by affixing the origin of the polar coordinate system for point $z_k$ at the point $z_{k-1}$ and measuring the polar angle $\phi_k$ {\it always with respect to the} x {\it axis}.  For example for $n=3$ the relevant cancellation occurs because $\Delta_2=\rho_2$ and $\Delta_3 = \rho_3$, and again $\phi_1$ can be integrated directly, resulting in (with $\Delta = \Delta_1 + \Delta_2 +\Delta_3 + \Delta_4 $)

\begin{equation}
    c_2 = -\frac{(2\pi)}{(4\pi)^{9/2} T^{3/2}}
    \int_{0}^{\rho_{max}}\!\!\!\!\!\!\!\! d\rho_1  \int_{0}^{\rho_{max}} \!\!\!\!\!\!\!\!d\rho_2  \int_{0}^{\rho_{max}}\!\!\!\!\!\!\!\! d\rho_3 
    \int_0^{2\pi} \!\!\!d\phi_2  \int_0^{2\pi}\!\!\! d\phi_3 \,\frac{e^{-\Delta^2/4T} \Delta \rho_1}{\sqrt{(\rho_1^2 + d^2)} \Delta_4}\,,
\end{equation}
where now

\begin{equation}
   \Delta_4= \sqrt{\rho_1^2 +\rho_2^2 +\rho_3^2 
    -2\rho_1 (\rho_2 \cos \phi_2 + \rho_3 \cos \phi_3) +2 \rho_2 \rho_3 
    \cos (\phi_2 - \phi_3)}\,.
\end{equation}

\subsection{Error Estimation} \label{Err}
 
 To assess the total error we made a simple estimate using the fact that the sequence of Pad\'e-Shanks approximations for all our cases of study in seen to be alternating around the analytical answer (this behaviour of course can be observed even if the analytical answer is unknown), a rough estimate is given by the Leibniz criterion as the difference of the last two approximations
\begin{equation}
    \epsilon (T)= |S_2 (T)- S_1 (T)|/|S_2 (T)|\,.
\end{equation}
The relative error given by $\epsilon$ goes from $0.7\%$, in the regions where it overestimates the actual deviation, to about $5\%$ in the regions where it is a good measure of the deviation; it stays about $3\%-4\%$ in a consistent manner throughout the intermediate regions. We can conclude that $\epsilon$ overestimates the main source of systematic error in our numerical calculations since the quadrature error $\epsilon_Q$ is usually much smaller. It is also worth noting that the precision in the results obtained by this method is mostly greater than the estimate $\epsilon$, as evidenced by Figs. \ref{SpherePlot1}, \ref{SpherePlot2} and \ref{PlanePlot} and is expected to improve if more coefficients $c_n$ are calculated and taken into account.

\section{Pad\'e approximants and Shanks transformation} \label{PShApp}

Pad\'e approximants are a sequence $\{P^M_N \}$ of  rational functions that approximate in a certain sense a given function $f(x)$. In the main text of this manuscript this function is the quantum mechanical kernel. These rational functions are written as the quotient of polynomials $ P^M_N (x):=P_M (x)/Q_N(x)$ of orders $M$ and $N$ respectively. If the function $f(x)$ is analytic and has a Taylor series $f(x) = \sum_{n=0}^{\infty} c_n x^n$ then the functions $P^M_N$ are constructed so that their Taylor series matches up to order $M+N$ with that of $f$, i.e.

\begin{equation}
    P^M_N (x) = c_0 + c_1 x + c_2 x^2 + \cdots + c_{M+N}x^{M+N} + O(x^{M+N+1})\,.
\end{equation}
This requirement, along with the normalisation 
\begin{equation}
    P^M_N (x) := \frac{\sum_{i=0}^{M}a_i x^i}{1 + \sum_{j=1}^{N} b_j x^j}\,,
\end{equation}
uniquely determines the coefficients $a_i$ and $b_j$ through a linear system of equations that can be solved using Cramer's rule; the result is well known in the literature \cite{bender2013advanced}

\begin{equation} \label{PadeMN}
   P^M_N (x) = \frac{ \left| \begin{array}{cccc}
        x^N \varphi _{M-N} & x^{N-1} \varphi_{M-N+1}&\cdots & \varphi_M  \\
        c_{M-N+1}& c_{M-N+2} &\cdots & c_{M+1} \\
        \vdots & \vdots & \vdots & \vdots \\
        c_M & c_{M+1} & \cdots & c_{M+N}
   \end{array} \right| }{\left| \begin{array}{cccc}
        x^N  & x^{N-1} &\cdots & 1  \\
        c_{M-N+1}& c_{M-N+2} &\cdots & c_{M+1} \\
        \vdots & \vdots & \vdots & \vdots \\
        c_M & c_{M+1} & \cdots & c_{M+N}
   \end{array} \right|} 
\end{equation}

\noindent where $\varphi_L := \sum_{p = 0}^L c_p x^p$. In the main text this method is applied to the Taylor expansion of the kernel given in (\ref{Khitseries}).

In general if we are looking for the value $f(\infty)$ the limit $\lim_{x\to \infty} P^{M}_N (x)$ will be zero if $N>M$ and divergent if $N<M$, but the idea was put forward in \cite{bender1994determination} that $M=N$ produces a finite limit that can be used to approximate $f(\infty)$ for a certain class of functions $f$. The ensuing subsequence of approximants $P^N_N (x)$ are called the diagonal Pad\'e approximants. This is the underpinning idea in proposing (\ref{Kpade}), whose strong coupling limit we estimate by a sequence of Pad\'e approximants determined in terms of integrals of the hit function.

\subsection{Shanks transformation}
Given an infinite sequence $\{ \alpha_n \}$ that converges to $\alpha = \lim_{n\to \infty} \alpha_n$ the convergence oftentimes behaves asymptotically in the large $n$ limit as 
\begin{equation} \label{asymp}
    \alpha_{n} \sim \alpha  + Z q^n,\qquad |q|<1,\quad \mbox{as }n\to \infty\,, 
\end{equation}
for some constants $Z,q$. Given this hypothesis the first order Shanks transformation \cite{bender2013advanced} consists of the solution for a numerical estimator $\mathcal{S}$ for the limiting value of the sequence $\alpha$, the idea is to replace $\sim$ by $=$ above and demand that the estimator $\mathcal{S}$ satisfies the linear system
\begin{eqnarray} \nonumber
\alpha_{n+1} & = & \mathcal{S} + Z q^{n+1} ,\\ 
\alpha_{n} & = & \mathcal{S} + Z q^n, \\ \nonumber
\alpha_{n-1} & = & \mathcal{S} + Z q^{n-1} .
\end{eqnarray}
The estimator found by solving this system is itself a new sequence $\{ \mathcal{S}(\alpha_n)\}$ that relates non-linearly to the original sequence $\{\alpha_n\}$  and is called the (first order) Shanks transformation of $\{\alpha_n\}$ 

\begin{equation} \label{Shanks}
    \mathcal{S}(\alpha_n) := \frac{\alpha_{n+1} \alpha_{n-1} -\alpha_{n}^2}{\alpha_{n+1} + \alpha_{n-1} -2\alpha_{n}}  \,.
\end{equation}
The Shanks transformation is useful because it often converges faster than the original sequence. When only a limited number of terms in the sequence are available it is often necessary to iterate the Shanks transformation to hasten convergence, for example $\mathcal{S}(\mathcal{S} (\alpha_n))$ is the second iteration. In the numerical calculations presented in section \ref{Benchmark} we have used the first and second iteration, $S_1$ and $S_2$, to show how the numerical results obtained are already significantly improved. We point out that to use the Shanks transformation in this manner we need to calculate the coefficients at least up to the term $\lambda^4$ in (\ref{Khitseries}). We should also point out that the validity of this method rests upon the hypothesis (\ref{asymp}) which can fail in some situations, it is nevertheless possible to lend support to the validity of this hypothesis if enough numerical data is available or if a theoretical arguments allow for its justification. Both, the Pad\'e extrapolation method and the Shanks transformation based acceleration of convergence are usually validated {\it post hoc} based on their rates of convergence and their numerical self consistency, many more details can be found e.g. in chapter 8 of \cite{bender2013advanced}.

\bibliography{refs.bib}

\begin{thebibliography}{52}%
\makeatletter
\providecommand \@ifxundefined [1]{%
 \@ifx{#1\undefined}
}%
\providecommand \@ifnum [1]{%
 \ifnum #1\expandafter \@firstoftwo
 \else \expandafter \@secondoftwo
 \fi
}%
\providecommand \@ifx [1]{%
 \ifx #1\expandafter \@firstoftwo
 \else \expandafter \@secondoftwo
 \fi
}%
\providecommand \natexlab [1]{#1}%
\providecommand \enquote  [1]{``#1''}%
\providecommand \bibnamefont  [1]{#1}%
\providecommand \bibfnamefont [1]{#1}%
\providecommand \citenamefont [1]{#1}%
\providecommand \href@noop [0]{\@secondoftwo}%
\providecommand \href [0]{\begingroup \@sanitize@url \@href}%
\providecommand \@href[1]{\@@startlink{#1}\@@href}%
\providecommand \@@href[1]{\endgroup#1\@@endlink}%
\providecommand \@sanitize@url [0]{\catcode `\\12\catcode `\$12\catcode
  `\&12\catcode `\#12\catcode `\^12\catcode `\_12\catcode `\%12\relax}%
\providecommand \@@startlink[1]{}%
\providecommand \@@endlink[0]{}%
\providecommand \url  [0]{\begingroup\@sanitize@url \@url }%
\providecommand \@url [1]{\endgroup\@href {#1}{\urlprefix }}%
\providecommand \urlprefix  [0]{URL }%
\providecommand \Eprint [0]{\href }%
\providecommand \doibase [0]{https://doi.org/}%
\providecommand \selectlanguage [0]{\@gobble}%
\providecommand \bibinfo  [0]{\@secondoftwo}%
\providecommand \bibfield  [0]{\@secondoftwo}%
\providecommand \translation [1]{[#1]}%
\providecommand \BibitemOpen [0]{}%
\providecommand \bibitemStop [0]{}%
\providecommand \bibitemNoStop [0]{.\EOS\space}%
\providecommand \EOS [0]{\spacefactor3000\relax}%
\providecommand \BibitemShut  [1]{\csname bibitem#1\endcsname}%
\let\auto@bib@innerbib\@empty
\bibitem [{\citenamefont {Grosche}(1990)}]{grosche1990path}%
  \BibitemOpen
  \bibfield  {author} {\bibinfo {author} {\bibfnamefont {C.}~\bibnamefont
  {Grosche}},\ }\bibfield  {title} {\bibinfo {title} {Path integrals for
  potential problems with $\delta$-function perturbation},\ }\href@noop {}
  {\bibfield  {journal} {\bibinfo  {journal} {Journal of Physics A:
  Mathematical and General}\ }\textbf {\bibinfo {volume} {23}},\ \bibinfo
  {pages} {5205} (\bibinfo {year} {1990})}\BibitemShut {NoStop}%
\bibitem [{\citenamefont {Grosche}(1993)}]{grosche1993delta}%
  \BibitemOpen
  \bibfield  {author} {\bibinfo {author} {\bibfnamefont {C.}~\bibnamefont
  {Grosche}},\ }\bibfield  {title} {\bibinfo {title} {$\delta$-function
  perturbations and boundary problems by path integration},\ }\href@noop {}
  {\bibfield  {journal} {\bibinfo  {journal} {Annalen der Physik}\ }\textbf
  {\bibinfo {volume} {505}},\ \bibinfo {pages} {557} (\bibinfo {year}
  {1993})}\BibitemShut {NoStop}%
\bibitem [{\citenamefont {Grosche}(1995)}]{grosche1995delta}%
  \BibitemOpen
  \bibfield  {author} {\bibinfo {author} {\bibfnamefont {C.}~\bibnamefont
  {Grosche}},\ }\bibfield  {title} {\bibinfo {title} {$\delta$'-function
  perturbations and neumann boundary conditions by path integration},\
  }\href@noop {} {\bibfield  {journal} {\bibinfo  {journal} {Journal of Physics
  A: Mathematical and General}\ }\textbf {\bibinfo {volume} {28}},\ \bibinfo
  {pages} {L99} (\bibinfo {year} {1995})}\BibitemShut {NoStop}%
\bibitem [{\citenamefont {Edwards}\ \emph {et~al.}(2018)\citenamefont
  {Edwards}, \citenamefont {Gerber}, \citenamefont {Schubert}, \citenamefont
  {Trejo},\ and\ \citenamefont {Weber}}]{PvHz}%
  \BibitemOpen
  \bibfield  {author} {\bibinfo {author} {\bibfnamefont {J.~P.}\ \bibnamefont
  {Edwards}}, \bibinfo {author} {\bibfnamefont {U.}~\bibnamefont {Gerber}},
  \bibinfo {author} {\bibfnamefont {C.}~\bibnamefont {Schubert}}, \bibinfo
  {author} {\bibfnamefont {M.~A.}\ \bibnamefont {Trejo}},\ and\ \bibinfo
  {author} {\bibfnamefont {A.}~\bibnamefont {Weber}},\ }\bibfield  {title}
  {\bibinfo {title} {{Integral transforms of the quantum mechanical path
  integral: hit function and path averaged potential}},\ }\href
  {https://doi.org/10.1103/PhysRevE.97.042114} {\bibfield  {journal} {\bibinfo
  {journal} {Phys. Rev. E}\ }\textbf {\bibinfo {volume} {97}},\ \bibinfo
  {pages} {042114} (\bibinfo {year} {2018})},\ \Eprint
  {https://arxiv.org/abs/1709.04984} {arXiv:1709.04984 [quant-ph]} \BibitemShut
  {NoStop}%
\bibitem [{\citenamefont {Jizba}\ and\ \citenamefont
  {Zatloukal}(2015)}]{PhysRevE.92.062137}%
  \BibitemOpen
  \bibfield  {author} {\bibinfo {author} {\bibfnamefont {P.}~\bibnamefont
  {Jizba}}\ and\ \bibinfo {author} {\bibfnamefont {V.}~\bibnamefont
  {Zatloukal}},\ }\bibfield  {title} {\bibinfo {title} {Local-time
  representation of path integrals},\ }\href
  {https://doi.org/10.1103/PhysRevE.92.062137} {\bibfield  {journal} {\bibinfo
  {journal} {Phys. Rev. E}\ }\textbf {\bibinfo {volume} {92}},\ \bibinfo
  {pages} {062137} (\bibinfo {year} {2015})}\BibitemShut {NoStop}%
\bibitem [{\citenamefont {Zatloukal}(2017)}]{PhysRevE.95.052136}%
  \BibitemOpen
  \bibfield  {author} {\bibinfo {author} {\bibfnamefont {V.}~\bibnamefont
  {Zatloukal}},\ }\bibfield  {title} {\bibinfo {title} {Local time of l\'evy
  random walks: A path integral approach},\ }\href
  {https://doi.org/10.1103/PhysRevE.95.052136} {\bibfield  {journal} {\bibinfo
  {journal} {Phys. Rev. E}\ }\textbf {\bibinfo {volume} {95}},\ \bibinfo
  {pages} {052136} (\bibinfo {year} {2017})}\BibitemShut {NoStop}%
\bibitem [{\citenamefont {Curry}\ and\ \citenamefont
  {Mansfield}(2018)}]{Curry:2017cnu}%
  \BibitemOpen
  \bibfield  {author} {\bibinfo {author} {\bibfnamefont {C.}~\bibnamefont
  {Curry}}\ and\ \bibinfo {author} {\bibfnamefont {P.}~\bibnamefont
  {Mansfield}},\ }\bibfield  {title} {\bibinfo {title} {{Intersection of
  world-lines on curved surfaces and path-ordering of the Wilson loop}},\
  }\href {https://doi.org/10.1007/JHEP06(2018)081} {\bibfield  {journal}
  {\bibinfo  {journal} {JHEP}\ }\textbf {\bibinfo {volume} {06}},\ \bibinfo
  {pages} {081}},\ \Eprint {https://arxiv.org/abs/1712.04760} {arXiv:1712.04760
  [hep-th]} \BibitemShut {NoStop}%
\bibitem [{\citenamefont {Edwards}(2016)}]{Edwards:2015hka}%
  \BibitemOpen
  \bibfield  {author} {\bibinfo {author} {\bibfnamefont {J.~P.}\ \bibnamefont
  {Edwards}},\ }\bibfield  {title} {\bibinfo {title} {{Contact interactions
  between particle worldlines}},\ }\href
  {https://doi.org/10.1007/JHEP01(2016)033} {\bibfield  {journal} {\bibinfo
  {journal} {JHEP}\ }\textbf {\bibinfo {volume} {01}},\ \bibinfo {pages}
  {033}},\ \Eprint {https://arxiv.org/abs/1506.08130} {arXiv:1506.08130
  [hep-th]} \BibitemShut {NoStop}%
\bibitem [{\citenamefont {Mansfield}(2012)}]{Mansfield:2011eq}%
  \BibitemOpen
  \bibfield  {author} {\bibinfo {author} {\bibfnamefont {P.}~\bibnamefont
  {Mansfield}},\ }\bibfield  {title} {\bibinfo {title} {{Faraday's lines of
  force as strings: from Gauss' law to the arrow of time}},\ }\href
  {https://doi.org/10.1007/JHEP10(2012)149} {\bibfield  {journal} {\bibinfo
  {journal} {JHEP}\ }\textbf {\bibinfo {volume} {10}},\ \bibinfo {pages}
  {149}},\ \Eprint {https://arxiv.org/abs/1108.5094} {arXiv:1108.5094 [hep-th]}
  \BibitemShut {NoStop}%
\bibitem [{\citenamefont {Kronig}\ and\ \citenamefont
  {Penney}(1931)}]{kronig1931quantum}%
  \BibitemOpen
  \bibfield  {author} {\bibinfo {author} {\bibfnamefont {R.~d.~L.}\
  \bibnamefont {Kronig}}\ and\ \bibinfo {author} {\bibfnamefont {W.~G.}\
  \bibnamefont {Penney}},\ }\bibfield  {title} {\bibinfo {title} {Quantum
  mechanics of electrons in crystal lattices},\ }\href@noop {} {\bibfield
  {journal} {\bibinfo  {journal} {Proceedings of the Royal Society of London.
  Series A, Containing Papers of a Mathematical and Physical Character}\
  }\textbf {\bibinfo {volume} {130}},\ \bibinfo {pages} {499} (\bibinfo {year}
  {1931})}\BibitemShut {NoStop}%
\bibitem [{\citenamefont {Demkov}\ and\ \citenamefont
  {Ostrovskii}(2013)}]{demkov2013zero}%
  \BibitemOpen
  \bibfield  {author} {\bibinfo {author} {\bibfnamefont {Y.~N.}\ \bibnamefont
  {Demkov}}\ and\ \bibinfo {author} {\bibfnamefont {V.~N.}\ \bibnamefont
  {Ostrovskii}},\ }\href@noop {} {\emph {\bibinfo {title} {Zero-range
  potentials and their applications in atomic physics}}}\ (\bibinfo
  {publisher} {Springer Science \& Business Media},\ \bibinfo {year}
  {2013})\BibitemShut {NoStop}%
\bibitem [{\citenamefont {Silva}\ \emph {et~al.}(2016)\citenamefont {Silva},
  \citenamefont {Braga},\ and\ \citenamefont {Alves}}]{PhysRevD.94.105009}%
  \BibitemOpen
  \bibfield  {author} {\bibinfo {author} {\bibfnamefont {J.~D.~L.}\
  \bibnamefont {Silva}}, \bibinfo {author} {\bibfnamefont {A.~N.}\ \bibnamefont
  {Braga}},\ and\ \bibinfo {author} {\bibfnamefont {D.~T.}\ \bibnamefont
  {Alves}},\ }\bibfield  {title} {\bibinfo {title} {Dynamical casimir effect
  with
  $\ensuremath{\delta}\ensuremath{-}{\ensuremath{\delta}}^{\ensuremath{'}}$
  mirrors},\ }\href {https://doi.org/10.1103/PhysRevD.94.105009} {\bibfield
  {journal} {\bibinfo  {journal} {Phys. Rev. D}\ }\textbf {\bibinfo {volume}
  {94}},\ \bibinfo {pages} {105009} (\bibinfo {year} {2016})}\BibitemShut
  {NoStop}%
\bibitem [{\citenamefont {Sch{\"a}fer}\ \emph {et~al.}(2016)\citenamefont
  {Sch{\"a}fer}, \citenamefont {Huet},\ and\ \citenamefont {Gies}}]{gies1}%
  \BibitemOpen
  \bibfield  {author} {\bibinfo {author} {\bibfnamefont {M.}~\bibnamefont
  {Sch{\"a}fer}}, \bibinfo {author} {\bibfnamefont {I.}~\bibnamefont {Huet}},\
  and\ \bibinfo {author} {\bibfnamefont {H.}~\bibnamefont {Gies}},\ }\bibfield
  {title} {\bibinfo {title} {Worldline numerics for energy-momentum tensors in
  casimir geometries},\ }\href@noop {} {\bibfield  {journal} {\bibinfo
  {journal} {Journal of Physics A: Mathematical and Theoretical}\ }\textbf
  {\bibinfo {volume} {49}},\ \bibinfo {pages} {135402} (\bibinfo {year}
  {2016})}\BibitemShut {NoStop}%
\bibitem [{\citenamefont {Gies}\ \emph {et~al.}(2003)\citenamefont {Gies},
  \citenamefont {Langfeld},\ and\ \citenamefont {Moyaerts}}]{gies2}%
  \BibitemOpen
  \bibfield  {author} {\bibinfo {author} {\bibfnamefont {H.}~\bibnamefont
  {Gies}}, \bibinfo {author} {\bibfnamefont {K.}~\bibnamefont {Langfeld}},\
  and\ \bibinfo {author} {\bibfnamefont {L.}~\bibnamefont {Moyaerts}},\
  }\bibfield  {title} {\bibinfo {title} {Casimir effect on the worldline},\
  }\href@noop {} {\bibfield  {journal} {\bibinfo  {journal} {Journal of High
  Energy Physics}\ }\textbf {\bibinfo {volume} {2003}},\ \bibinfo {pages} {018}
  (\bibinfo {year} {2003})}\BibitemShut {NoStop}%
\bibitem [{\citenamefont {Gies}\ and\ \citenamefont
  {Klingm{\"u}ller}(2006)}]{gies3}%
  \BibitemOpen
  \bibfield  {author} {\bibinfo {author} {\bibfnamefont {H.}~\bibnamefont
  {Gies}}\ and\ \bibinfo {author} {\bibfnamefont {K.}~\bibnamefont
  {Klingm{\"u}ller}},\ }\bibfield  {title} {\bibinfo {title} {Worldline
  algorithms for casimir configurations},\ }\href@noop {} {\bibfield  {journal}
  {\bibinfo  {journal} {Physical Review D}\ }\textbf {\bibinfo {volume} {74}},\
  \bibinfo {pages} {045002} (\bibinfo {year} {2006})}\BibitemShut {NoStop}%
\bibitem [{\citenamefont {Clark}\ \emph {et~al.}(1980)\citenamefont {Clark},
  \citenamefont {Menikoff},\ and\ \citenamefont {Sharp}}]{Clark:1980xt}%
  \BibitemOpen
  \bibfield  {author} {\bibinfo {author} {\bibfnamefont {T.~E.}\ \bibnamefont
  {Clark}}, \bibinfo {author} {\bibfnamefont {R.}~\bibnamefont {Menikoff}},\
  and\ \bibinfo {author} {\bibfnamefont {D.~H.}\ \bibnamefont {Sharp}},\
  }\bibfield  {title} {\bibinfo {title} {{Quantum mechanics on the half line
  using path integrals}},\ }\href {https://doi.org/10.1103/PhysRevD.22.3012}
  {\bibfield  {journal} {\bibinfo  {journal} {Phys. Rev. D}\ }\textbf {\bibinfo
  {volume} {22}},\ \bibinfo {pages} {3012} (\bibinfo {year}
  {1980})}\BibitemShut {NoStop}%
\bibitem [{\citenamefont {Farhi}\ and\ \citenamefont
  {Gutmann}(1990)}]{Farhi:1989jz}%
  \BibitemOpen
  \bibfield  {author} {\bibinfo {author} {\bibfnamefont {E.}~\bibnamefont
  {Farhi}}\ and\ \bibinfo {author} {\bibfnamefont {S.}~\bibnamefont
  {Gutmann}},\ }\bibfield  {title} {\bibinfo {title} {{The functional integral
  on the half line}},\ }\href {https://doi.org/10.1142/S0217751X90001422}
  {\bibfield  {journal} {\bibinfo  {journal} {Int. J. Mod. Phys. A}\ }\textbf
  {\bibinfo {volume} {5}},\ \bibinfo {pages} {3029} (\bibinfo {year}
  {1990})}\BibitemShut {NoStop}%
\bibitem [{\citenamefont {Bastianelli}\ \emph {et~al.}(2007)\citenamefont
  {Bastianelli}, \citenamefont {Corradini},\ and\ \citenamefont
  {Pisani}}]{Bastianelli:2006hq}%
  \BibitemOpen
  \bibfield  {author} {\bibinfo {author} {\bibfnamefont {F.}~\bibnamefont
  {Bastianelli}}, \bibinfo {author} {\bibfnamefont {O.}~\bibnamefont
  {Corradini}},\ and\ \bibinfo {author} {\bibfnamefont {P.~A.~G.}\ \bibnamefont
  {Pisani}},\ }\bibfield  {title} {\bibinfo {title} {{Worldline approach to
  quantum field theories on flat manifolds with boundaries}},\ }\href
  {https://doi.org/10.1088/1126-6708/2007/02/059} {\bibfield  {journal}
  {\bibinfo  {journal} {JHEP}\ }\textbf {\bibinfo {volume} {02}},\ \bibinfo
  {pages} {059}},\ \Eprint {https://arxiv.org/abs/hep-th/0612236}
  {arXiv:hep-th/0612236} \BibitemShut {NoStop}%
\bibitem [{\citenamefont {Bastianelli}\ \emph
  {et~al.}(2008{\natexlab{a}})\citenamefont {Bastianelli}, \citenamefont
  {Corradini},\ and\ \citenamefont {Pisani}}]{Bastianelli:2007jr}%
  \BibitemOpen
  \bibfield  {author} {\bibinfo {author} {\bibfnamefont {F.}~\bibnamefont
  {Bastianelli}}, \bibinfo {author} {\bibfnamefont {O.}~\bibnamefont
  {Corradini}},\ and\ \bibinfo {author} {\bibfnamefont {P.~A.~G.}\ \bibnamefont
  {Pisani}},\ }\bibfield  {title} {\bibinfo {title} {{Scalar field with Robin
  boundary conditions in the worldline formalism}},\ }\href
  {https://doi.org/10.1088/1751-8113/41/16/164010} {\bibfield  {journal}
  {\bibinfo  {journal} {J. Phys. A}\ }\textbf {\bibinfo {volume} {41}},\
  \bibinfo {pages} {164010} (\bibinfo {year} {2008}{\natexlab{a}})},\ \Eprint
  {https://arxiv.org/abs/0710.4026} {arXiv:0710.4026 [hep-th]} \BibitemShut
  {NoStop}%
\bibitem [{\citenamefont {Bastianelli}\ \emph
  {et~al.}(2008{\natexlab{b}})\citenamefont {Bastianelli}, \citenamefont
  {Corradini}, \citenamefont {Pisani},\ and\ \citenamefont
  {Schubert}}]{bastianelli2008scalar}%
  \BibitemOpen
  \bibfield  {author} {\bibinfo {author} {\bibfnamefont {F.}~\bibnamefont
  {Bastianelli}}, \bibinfo {author} {\bibfnamefont {O.}~\bibnamefont
  {Corradini}}, \bibinfo {author} {\bibfnamefont {P.~A.}\ \bibnamefont
  {Pisani}},\ and\ \bibinfo {author} {\bibfnamefont {C.}~\bibnamefont
  {Schubert}},\ }\bibfield  {title} {\bibinfo {title} {Scalar heat kernel with
  boundary in the worldline formalism},\ }\href@noop {} {\bibfield  {journal}
  {\bibinfo  {journal} {Journal of High Energy Physics}\ }\textbf {\bibinfo
  {volume} {2008}},\ \bibinfo {pages} {095} (\bibinfo {year}
  {2008}{\natexlab{b}})}\BibitemShut {NoStop}%
\bibitem [{\citenamefont {Vinas}\ and\ \citenamefont
  {Pisani}(2011)}]{Vinas:2010ix}%
  \BibitemOpen
  \bibfield  {author} {\bibinfo {author} {\bibfnamefont {S.~A.~F.}\
  \bibnamefont {Vinas}}\ and\ \bibinfo {author} {\bibfnamefont {P.~A.~G.}\
  \bibnamefont {Pisani}},\ }\bibfield  {title} {\bibinfo {title}
  {{Semi-transparent boundary conditions in the worldline formalism}},\ }\href
  {https://doi.org/10.1088/1751-8113/44/29/295401} {\bibfield  {journal}
  {\bibinfo  {journal} {J. Phys. A}\ }\textbf {\bibinfo {volume} {44}},\
  \bibinfo {pages} {295401} (\bibinfo {year} {2011})},\ \Eprint
  {https://arxiv.org/abs/1012.2883} {arXiv:1012.2883 [hep-th]} \BibitemShut
  {NoStop}%
\bibitem [{\citenamefont {Corradini}\ \emph
  {et~al.}(2019{\natexlab{a}})\citenamefont {Corradini}, \citenamefont
  {Edwards}, \citenamefont {Huet}, \citenamefont {Manzo},\ and\ \citenamefont
  {Pisani}}]{Corradini:2019nbb}%
  \BibitemOpen
  \bibfield  {author} {\bibinfo {author} {\bibfnamefont {O.}~\bibnamefont
  {Corradini}}, \bibinfo {author} {\bibfnamefont {J.~P.}\ \bibnamefont
  {Edwards}}, \bibinfo {author} {\bibfnamefont {I.}~\bibnamefont {Huet}},
  \bibinfo {author} {\bibfnamefont {L.}~\bibnamefont {Manzo}},\ and\ \bibinfo
  {author} {\bibfnamefont {P.}~\bibnamefont {Pisani}},\ }\bibfield  {title}
  {\bibinfo {title} {{Worldline formalism for a confined scalar field}},\
  }\href {https://doi.org/10.1007/JHEP08(2019)037} {\bibfield  {journal}
  {\bibinfo  {journal} {JHEP}\ }\textbf {\bibinfo {volume} {08}},\ \bibinfo
  {pages} {037}},\ \Eprint {https://arxiv.org/abs/1905.00945} {arXiv:1905.00945
  [hep-th]} \BibitemShut {NoStop}%
\bibitem [{\citenamefont {Carreau}(1992)}]{Carreau:1991yx}%
  \BibitemOpen
  \bibfield  {author} {\bibinfo {author} {\bibfnamefont {M.}~\bibnamefont
  {Carreau}},\ }\bibfield  {title} {\bibinfo {title} {{The functional integral
  for a free particle on a half plane}},\ }\href
  {https://doi.org/10.1063/1.529812} {\bibfield  {journal} {\bibinfo  {journal}
  {J. Math. Phys.}\ }\textbf {\bibinfo {volume} {33}},\ \bibinfo {pages} {4139}
  (\bibinfo {year} {1992})},\ \Eprint {https://arxiv.org/abs/hep-th/9208052}
  {arXiv:hep-th/9208052} \BibitemShut {NoStop}%
\bibitem [{\citenamefont {Asorey}\ \emph {et~al.}(2006)\citenamefont {Asorey},
  \citenamefont {Ibort},\ and\ \citenamefont {Marmo}}]{Asorey:2006ij}%
  \BibitemOpen
  \bibfield  {author} {\bibinfo {author} {\bibfnamefont {M.}~\bibnamefont
  {Asorey}}, \bibinfo {author} {\bibfnamefont {A.}~\bibnamefont {Ibort}},\ and\
  \bibinfo {author} {\bibfnamefont {G.}~\bibnamefont {Marmo}},\ }\bibfield
  {title} {\bibinfo {title} {{Path integrals and boundary conditions}},\ }in\
  \href@noop {} {\emph {\bibinfo {booktitle} {{32nd International Meeting on
  Fundamental Physics}}}}\ (\bibinfo {year} {2006})\ \Eprint
  {https://arxiv.org/abs/quant-ph/0609023} {arXiv:quant-ph/0609023}
  \BibitemShut {NoStop}%
\bibitem [{\citenamefont {Asorey}\ \emph {et~al.}(2007)\citenamefont {Asorey},
  \citenamefont {Munoz-Castaneda},\ and\ \citenamefont
  {Clemente-Gallardo}}]{Asorey:2007zza}%
  \BibitemOpen
  \bibfield  {author} {\bibinfo {author} {\bibfnamefont {M.}~\bibnamefont
  {Asorey}}, \bibinfo {author} {\bibfnamefont {J.~M.}\ \bibnamefont
  {Munoz-Castaneda}},\ and\ \bibinfo {author} {\bibfnamefont {J.}~\bibnamefont
  {Clemente-Gallardo}},\ }\bibfield  {title} {\bibinfo {title} {{Boundary
  conditions: The path integral approach}},\ }\href
  {https://doi.org/10.1088/1742-6596/87/1/012004} {\bibfield  {journal}
  {\bibinfo  {journal} {J. Phys. Conf. Ser.}\ }\textbf {\bibinfo {volume}
  {87}},\ \bibinfo {pages} {012004} (\bibinfo {year} {2007})},\ \Eprint
  {https://arxiv.org/abs/0712.4353} {arXiv:0712.4353 [quant-ph]} \BibitemShut
  {NoStop}%
\bibitem [{\citenamefont {Auerbach}\ \emph {et~al.}(1984)\citenamefont
  {Auerbach}, \citenamefont {Kivelson},\ and\ \citenamefont
  {Nicole}}]{PhysRevLett.53.411}%
  \BibitemOpen
  \bibfield  {author} {\bibinfo {author} {\bibfnamefont {A.}~\bibnamefont
  {Auerbach}}, \bibinfo {author} {\bibfnamefont {S.}~\bibnamefont {Kivelson}},\
  and\ \bibinfo {author} {\bibfnamefont {D.}~\bibnamefont {Nicole}},\
  }\bibfield  {title} {\bibinfo {title} {Path decomposition for
  multidimensional tunneling},\ }\href
  {https://doi.org/10.1103/PhysRevLett.53.411} {\bibfield  {journal} {\bibinfo
  {journal} {Phys. Rev. Lett.}\ }\textbf {\bibinfo {volume} {53}},\ \bibinfo
  {pages} {411} (\bibinfo {year} {1984})}\BibitemShut {NoStop}%
\bibitem [{\citenamefont {Halliwell}(1995)}]{Halliwell:1995jh}%
  \BibitemOpen
  \bibfield  {author} {\bibinfo {author} {\bibfnamefont {J.~J.}\ \bibnamefont
  {Halliwell}},\ }\bibfield  {title} {\bibinfo {title} {{An operator derivation
  of the path decomposition expansion}},\ }\href
  {https://doi.org/10.1016/0375-9601(95)00703-6} {\bibfield  {journal}
  {\bibinfo  {journal} {Phys. Lett. A}\ }\textbf {\bibinfo {volume} {207}},\
  \bibinfo {pages} {237} (\bibinfo {year} {1995})},\ \Eprint
  {https://arxiv.org/abs/quant-ph/9506021} {arXiv:quant-ph/9506021}
  \BibitemShut {NoStop}%
\bibitem [{\citenamefont {Yearsley}(2009)}]{Yearsley_2009}%
  \BibitemOpen
  \bibfield  {author} {\bibinfo {author} {\bibfnamefont {J.~M.}\ \bibnamefont
  {Yearsley}},\ }\bibfield  {title} {\bibinfo {title} {The propagator for the
  step potential using the path decomposition expansion},\ }\href
  {https://doi.org/10.1088/1742-6596/174/1/012072} {\bibfield  {journal}
  {\bibinfo  {journal} {Journal of Physics: Conference Series}\ }\textbf
  {\bibinfo {volume} {174}},\ \bibinfo {pages} {012072} (\bibinfo {year}
  {2009})}\BibitemShut {NoStop}%
\bibitem [{\citenamefont {Yearsley}(2008)}]{Yearsley_2008}%
  \BibitemOpen
  \bibfield  {author} {\bibinfo {author} {\bibfnamefont {J.~M.}\ \bibnamefont
  {Yearsley}},\ }\bibfield  {title} {\bibinfo {title} {The propagator for the
  step potential and delta function potential using the path decomposition
  expansion},\ }\href {https://doi.org/10.1088/1751-8113/41/28/285301}
  {\bibfield  {journal} {\bibinfo  {journal} {Journal of Physics A:
  Mathematical and Theoretical}\ }\textbf {\bibinfo {volume} {41}},\ \bibinfo
  {pages} {285301} (\bibinfo {year} {2008})}\BibitemShut {NoStop}%
\bibitem [{\citenamefont {Koch}\ and\ \citenamefont
  {Mu\~noz}(2020)}]{Koch:2020dql}%
  \BibitemOpen
  \bibfield  {author} {\bibinfo {author} {\bibfnamefont {B.}~\bibnamefont
  {Koch}}\ and\ \bibinfo {author} {\bibfnamefont {E.}~\bibnamefont {Mu\~noz}},\
  }\bibfield  {title} {\bibinfo {title} {{Path integral of the relativistic
  point particle in Minkowski space}},\ }\href@noop {} {\  (\bibinfo {year}
  {2020})},\ \Eprint {https://arxiv.org/abs/2012.15242} {arXiv:2012.15242
  [hep-th]} \BibitemShut {NoStop}%
\bibitem [{\citenamefont {Koch}\ and\ \citenamefont
  {Mu\~noz}(2018)}]{Koch:2017nha}%
  \BibitemOpen
  \bibfield  {author} {\bibinfo {author} {\bibfnamefont {B.}~\bibnamefont
  {Koch}}\ and\ \bibinfo {author} {\bibfnamefont {E.}~\bibnamefont {Mu\~noz}},\
  }\bibfield  {title} {\bibinfo {title} {{The stepwise path integral of the
  relativistic point particle}},\ }\href
  {https://doi.org/10.1140/epjc/s10052-018-5753-9} {\bibfield  {journal}
  {\bibinfo  {journal} {Eur. Phys. J. C}\ }\textbf {\bibinfo {volume} {78}},\
  \bibinfo {pages} {278} (\bibinfo {year} {2018})},\ \Eprint
  {https://arxiv.org/abs/1706.05388} {arXiv:1706.05388 [hep-th]} \BibitemShut
  {NoStop}%
\bibitem [{\citenamefont {Franchino-Vi\~nas}\ and\ \citenamefont
  {Mazzitelli}(2021)}]{Franchino-Vinas:2020okl}%
  \BibitemOpen
  \bibfield  {author} {\bibinfo {author} {\bibfnamefont {S.~A.}\ \bibnamefont
  {Franchino-Vi\~nas}}\ and\ \bibinfo {author} {\bibfnamefont {F.~D.}\
  \bibnamefont {Mazzitelli}},\ }\bibfield  {title} {\bibinfo {title}
  {{Effective action for delta potentials: spacetime-dependent inhomogeneities
  and Casimir self-energy}},\ }\href
  {https://doi.org/10.1103/PhysRevD.103.065006} {\bibfield  {journal} {\bibinfo
   {journal} {Phys. Rev. D}\ }\textbf {\bibinfo {volume} {103}},\ \bibinfo
  {pages} {065006} (\bibinfo {year} {2021})},\ \Eprint
  {https://arxiv.org/abs/2010.11144} {arXiv:2010.11144 [hep-th]} \BibitemShut
  {NoStop}%
\bibitem [{\citenamefont {Polyakov}(1987)}]{Polyakov:1987ez}%
  \BibitemOpen
  \bibfield  {author} {\bibinfo {author} {\bibfnamefont {A.~M.}\ \bibnamefont
  {Polyakov}},\ }\href@noop {} {\emph {\bibinfo {title} {{Gauge fields and
  strings}}}},\ Vol.~\bibinfo {volume} {3}\ (\bibinfo {year}
  {1987})\BibitemShut {NoStop}%
\bibitem [{\citenamefont {Feynman}\ \emph {et~al.}(2010)\citenamefont
  {Feynman}, \citenamefont {Hibbs},\ and\ \citenamefont
  {Styer}}]{feynman2010quantum}%
  \BibitemOpen
  \bibfield  {author} {\bibinfo {author} {\bibfnamefont {R.~P.}\ \bibnamefont
  {Feynman}}, \bibinfo {author} {\bibfnamefont {A.~R.}\ \bibnamefont {Hibbs}},\
  and\ \bibinfo {author} {\bibfnamefont {D.~F.}\ \bibnamefont {Styer}},\
  }\href@noop {} {\emph {\bibinfo {title} {Quantum mechanics and path
  integrals}}}\ (\bibinfo  {publisher} {Courier Corporation},\ \bibinfo {year}
  {2010})\BibitemShut {NoStop}%
\bibitem [{\citenamefont {Edwards}\ \emph {et~al.}(2019)\citenamefont
  {Edwards}, \citenamefont {Gerber}, \citenamefont {Schubert}, \citenamefont
  {Trejo}, \citenamefont {Tsiftsi},\ and\ \citenamefont
  {Weber}}]{Edwards:2019fjh}%
  \BibitemOpen
  \bibfield  {author} {\bibinfo {author} {\bibfnamefont {J.~P.}\ \bibnamefont
  {Edwards}}, \bibinfo {author} {\bibfnamefont {U.}~\bibnamefont {Gerber}},
  \bibinfo {author} {\bibfnamefont {C.}~\bibnamefont {Schubert}}, \bibinfo
  {author} {\bibfnamefont {M.~A.}\ \bibnamefont {Trejo}}, \bibinfo {author}
  {\bibfnamefont {T.}~\bibnamefont {Tsiftsi}},\ and\ \bibinfo {author}
  {\bibfnamefont {A.}~\bibnamefont {Weber}},\ }\bibfield  {title} {\bibinfo
  {title} {{Applications of the worldline Monte Carlo formalism in quantum
  mechanics}},\ }\href {https://doi.org/10.1016/j.aop.2019.167966} {\bibfield
  {journal} {\bibinfo  {journal} {Annals Phys.}\ }\textbf {\bibinfo {volume}
  {411}},\ \bibinfo {pages} {167966} (\bibinfo {year} {2019})},\ \Eprint
  {https://arxiv.org/abs/1903.00536} {arXiv:1903.00536 [quant-ph]} \BibitemShut
  {NoStop}%
\bibitem [{\citenamefont {Corradini}\ and\ \citenamefont
  {Muratori}(2020)}]{Corradini:2020tgk}%
  \BibitemOpen
  \bibfield  {author} {\bibinfo {author} {\bibfnamefont {O.}~\bibnamefont
  {Corradini}}\ and\ \bibinfo {author} {\bibfnamefont {M.}~\bibnamefont
  {Muratori}},\ }\bibfield  {title} {\bibinfo {title} {{A Monte Carlo approach
  to the worldline formalism in curved space}},\ }\href
  {https://doi.org/10.1007/JHEP11(2020)169} {\bibfield  {journal} {\bibinfo
  {journal} {JHEP}\ }\textbf {\bibinfo {volume} {11}},\ \bibinfo {pages}
  {169}},\ \Eprint {https://arxiv.org/abs/2006.02911} {arXiv:2006.02911
  [hep-th]} \BibitemShut {NoStop}%
\bibitem [{\citenamefont {Franchino-Vi\~nas}\ and\ \citenamefont
  {Gies}(2019)}]{Franchino-Vinas:2019udt}%
  \BibitemOpen
  \bibfield  {author} {\bibinfo {author} {\bibfnamefont {S.}~\bibnamefont
  {Franchino-Vi\~nas}}\ and\ \bibinfo {author} {\bibfnamefont {H.}~\bibnamefont
  {Gies}},\ }\bibfield  {title} {\bibinfo {title} {{Propagator from
  nonperturbative worldline dynamics}},\ }\href
  {https://doi.org/10.1103/PhysRevD.100.105020} {\bibfield  {journal} {\bibinfo
   {journal} {Phys. Rev. D}\ }\textbf {\bibinfo {volume} {100}},\ \bibinfo
  {pages} {105020} (\bibinfo {year} {2019})},\ \Eprint
  {https://arxiv.org/abs/1908.04532} {arXiv:1908.04532 [hep-th]} \BibitemShut
  {NoStop}%
\bibitem [{\citenamefont {Schubert}(2001)}]{ChrisRev}%
  \BibitemOpen
  \bibfield  {author} {\bibinfo {author} {\bibfnamefont {C.}~\bibnamefont
  {Schubert}},\ }\bibfield  {title} {\bibinfo {title} {{Perturbative quantum
  field theory in the string inspired formalism}},\ }\href
  {https://doi.org/10.1016/S0370-1573(01)00013-8} {\bibfield  {journal}
  {\bibinfo  {journal} {Phys. Rept.}\ }\textbf {\bibinfo {volume} {355}},\
  \bibinfo {pages} {73} (\bibinfo {year} {2001})},\ \Eprint
  {https://arxiv.org/abs/hep-th/0101036} {arXiv:hep-th/0101036} \BibitemShut
  {NoStop}%
\bibitem [{\citenamefont {Edwards}\ and\ \citenamefont
  {Schubert}(2019)}]{UsRep}%
  \BibitemOpen
  \bibfield  {author} {\bibinfo {author} {\bibfnamefont {J.~P.}\ \bibnamefont
  {Edwards}}\ and\ \bibinfo {author} {\bibfnamefont {C.}~\bibnamefont
  {Schubert}},\ }\bibfield  {title} {\bibinfo {title} {{Quantum mechanical path
  integrals in the first quantised approach to quantum field theory}}\
  }(\bibinfo {year} {2019})\ \Eprint {https://arxiv.org/abs/1912.10004}
  {arXiv:1912.10004 [hep-th]} \BibitemShut {NoStop}%
\bibitem [{\citenamefont {Schmidt}\ and\ \citenamefont
  {Schubert}(1998)}]{SchmidtRev}%
  \BibitemOpen
  \bibfield  {author} {\bibinfo {author} {\bibfnamefont {M.~G.}\ \bibnamefont
  {Schmidt}}\ and\ \bibinfo {author} {\bibfnamefont {C.}~\bibnamefont
  {Schubert}},\ }\bibfield  {title} {\bibinfo {title} {{Worldline path
  integrals as a calculational tool in quantum field theory}},\ }in\ \href@noop
  {} {\emph {\bibinfo {booktitle} {{6th International Conference on Path
  Integrals from PeV to TeV: 50 years from Feynman's Paper (PI 98)}}}}\
  (\bibinfo {year} {1998})\ \Eprint {https://arxiv.org/abs/hep-th/9810161}
  {arXiv:hep-th/9810161} \BibitemShut {NoStop}%
\bibitem [{\citenamefont {Corradini}\ \emph
  {et~al.}(2019{\natexlab{b}})\citenamefont {Corradini}, \citenamefont
  {Edwards}, \citenamefont {Huet}, \citenamefont {Manzo},\ and\ \citenamefont
  {Pisani}}]{pisanietal}%
  \BibitemOpen
  \bibfield  {author} {\bibinfo {author} {\bibfnamefont {O.}~\bibnamefont
  {Corradini}}, \bibinfo {author} {\bibfnamefont {J.~P.}\ \bibnamefont
  {Edwards}}, \bibinfo {author} {\bibfnamefont {I.}~\bibnamefont {Huet}},
  \bibinfo {author} {\bibfnamefont {L.}~\bibnamefont {Manzo}},\ and\ \bibinfo
  {author} {\bibfnamefont {P.}~\bibnamefont {Pisani}},\ }\bibfield  {title}
  {\bibinfo {title} {Worldline formalism for a confined scalar field},\
  }\href@noop {} {\bibfield  {journal} {\bibinfo  {journal} {Journal of High
  Energy Physics}\ }\textbf {\bibinfo {volume} {2019}},\ \bibinfo {pages} {1}
  (\bibinfo {year} {2019}{\natexlab{b}})}\BibitemShut {NoStop}%
\bibitem [{\citenamefont {Vassilevich}(2003)}]{vassilevich2003heat}%
  \BibitemOpen
  \bibfield  {author} {\bibinfo {author} {\bibfnamefont {D.~V.}\ \bibnamefont
  {Vassilevich}},\ }\bibfield  {title} {\bibinfo {title} {Heat kernel
  expansion: user's manual},\ }\href@noop {} {\bibfield  {journal} {\bibinfo
  {journal} {Physics reports}\ }\textbf {\bibinfo {volume} {388}},\ \bibinfo
  {pages} {279} (\bibinfo {year} {2003})}\BibitemShut {NoStop}%
\bibitem [{\citenamefont {Bender}\ and\ \citenamefont
  {Boettcher}(1994)}]{bender1994determination}%
  \BibitemOpen
  \bibfield  {author} {\bibinfo {author} {\bibfnamefont {C.~M.}\ \bibnamefont
  {Bender}}\ and\ \bibinfo {author} {\bibfnamefont {S.}~\bibnamefont
  {Boettcher}},\ }\bibfield  {title} {\bibinfo {title} {Determination of $f
  (\infty)$ from the asymptotic series for $f (x)$ about $x= 0$},\ }\href@noop
  {} {\bibfield  {journal} {\bibinfo  {journal} {Journal of Mathematical
  Physics}\ }\textbf {\bibinfo {volume} {35}},\ \bibinfo {pages} {1914}
  (\bibinfo {year} {1994})}\BibitemShut {NoStop}%
\bibitem [{\citenamefont {Gel'fand}\ and\ \citenamefont
  {Yaglom}(1960)}]{gelfand}%
  \BibitemOpen
  \bibfield  {author} {\bibinfo {author} {\bibfnamefont {I.~M.}\ \bibnamefont
  {Gel'fand}}\ and\ \bibinfo {author} {\bibfnamefont {A.~M.}\ \bibnamefont
  {Yaglom}},\ }\bibfield  {title} {\bibinfo {title} {Integration in functional
  spaces and its applications in quantum physics},\ }\href@noop {} {\bibfield
  {journal} {\bibinfo  {journal} {Journal of Mathematical Physics}\ }\textbf
  {\bibinfo {volume} {1}},\ \bibinfo {pages} {48} (\bibinfo {year}
  {1960})}\BibitemShut {NoStop}%
\bibitem [{\citenamefont {Gradshteyn}\ and\ \citenamefont
  {Ryzhik}(2007)}]{gradshteyn2007table}%
  \BibitemOpen
  \bibfield  {author} {\bibinfo {author} {\bibfnamefont {I.~S.}\ \bibnamefont
  {Gradshteyn}}\ and\ \bibinfo {author} {\bibfnamefont {I.~M.}\ \bibnamefont
  {Ryzhik}},\ }\href@noop {} {\emph {\bibinfo {title} {Table of integrals,
  series, and products}}}\ (\bibinfo  {publisher} {Academic press},\ \bibinfo
  {year} {2007})\BibitemShut {NoStop}%
\bibitem [{\citenamefont {Kleinert}(2004)}]{KleinertBook}%
  \BibitemOpen
  \bibfield  {author} {\bibinfo {author} {\bibfnamefont {H.}~\bibnamefont
  {Kleinert}},\ }\bibfield  {title} {\bibinfo {title} {{Path integrals in
  quantum mechanics, statistics, polymer physics, and financial markets}},\
  }\href@noop {} {\  (\bibinfo {year} {2004})}\BibitemShut {NoStop}%
\bibitem [{\citenamefont {Abramowitz}\ and\ \citenamefont
  {Stegun}(1964)}]{AbSt}%
  \BibitemOpen
  \bibinfo {editor} {\bibfnamefont {M.}~\bibnamefont {Abramowitz}}\ and\
  \bibinfo {editor} {\bibfnamefont {I.}~\bibnamefont {Stegun}},\ eds.,\
  \href@noop {} {\emph {\bibinfo {title} {Handbook of mathematical functions
  with formulas, graphs, and mathematical tables.}}},\ Vol.~\bibinfo {volume}
  {55}\ (\bibinfo  {publisher} {NBS},\ \bibinfo {year} {1964})\BibitemShut
  {NoStop}%
\bibitem [{{\relax DLMF} 7.18(i), dlmf.nist.gov/7.18.i()}]{DLMF}%
  \BibitemOpen
  {\relax DLMF} 7.18(i), dlmf.nist.gov/7.18.i,\ \href
  {https://dlmf.nist.gov/7.18.i} {\bibinfo {title} {{\it NIST Digital library
  of mathematical functions}}},\ \bibinfo {howpublished}
  {http://dlmf.nist.gov/, Release 1.1.2 of 2021-06-15},\ \bibinfo {note}
  {{\relax F.~W.~J. Olver, A.~B. {Olde Daalhuis}, D.~W. Lozier, B.~I.
  Schneider, R.~F. Boisvert, C.~W. Clark, B.~R. Miller, B.~V. Saunders, H.~S.
  Cohl, and M.~A. McClain, eds.}}\BibitemShut {Stop}%
\bibitem [{\citenamefont {Bender}\ and\ \citenamefont
  {Orszag}(2013)}]{bender2013advanced}%
  \BibitemOpen
  \bibfield  {author} {\bibinfo {author} {\bibfnamefont {C.~M.}\ \bibnamefont
  {Bender}}\ and\ \bibinfo {author} {\bibfnamefont {S.~A.}\ \bibnamefont
  {Orszag}},\ }\href@noop {} {\emph {\bibinfo {title} {Advanced mathematical
  methods for scientists and engineers}}}\ (\bibinfo  {publisher} {Springer
  Science \& Business Media},\ \bibinfo {year} {2013})\BibitemShut {NoStop}%
\bibitem [{\citenamefont {Bastianelli}\ \emph {et~al.}(2014)\citenamefont
  {Bastianelli}, \citenamefont {Huet}, \citenamefont {Schubert}, \citenamefont
  {Thakur},\ and\ \citenamefont {Weber}}]{Bastianelli:2014bfa}%
  \BibitemOpen
  \bibfield  {author} {\bibinfo {author} {\bibfnamefont {F.}~\bibnamefont
  {Bastianelli}}, \bibinfo {author} {\bibfnamefont {A.}~\bibnamefont {Huet}},
  \bibinfo {author} {\bibfnamefont {C.}~\bibnamefont {Schubert}}, \bibinfo
  {author} {\bibfnamefont {R.}~\bibnamefont {Thakur}},\ and\ \bibinfo {author}
  {\bibfnamefont {A.}~\bibnamefont {Weber}},\ }\bibfield  {title} {\bibinfo
  {title} {{Integral representations combining ladders and crossed-ladders}},\
  }\href {https://doi.org/10.1007/JHEP07(2014)066} {\bibfield  {journal}
  {\bibinfo  {journal} {JHEP}\ }\textbf {\bibinfo {volume} {07}},\ \bibinfo
  {pages} {066}},\ \Eprint {https://arxiv.org/abs/1405.7770} {arXiv:1405.7770
  [hep-ph]} \BibitemShut {NoStop}%
\bibitem [{\citenamefont {Fujiwara}\ \emph {et~al.}(1982)\citenamefont
  {Fujiwara}, \citenamefont {Osborn},\ and\ \citenamefont
  {Wilk}}]{Fujiwara:1981rf}%
  \BibitemOpen
  \bibfield  {author} {\bibinfo {author} {\bibfnamefont {Y.}~\bibnamefont
  {Fujiwara}}, \bibinfo {author} {\bibfnamefont {T.~A.}\ \bibnamefont
  {Osborn}},\ and\ \bibinfo {author} {\bibfnamefont {S.~F.~J.}\ \bibnamefont
  {Wilk}},\ }\bibfield  {title} {\bibinfo {title} {{Wigner-Kirkwood
  expansions}},\ }\href {https://doi.org/10.1103/PhysRevA.25.14} {\bibfield
  {journal} {\bibinfo  {journal} {Phys. Rev. A}\ }\textbf {\bibinfo {volume}
  {25}},\ \bibinfo {pages} {14} (\bibinfo {year} {1982})}\BibitemShut {NoStop}%
\bibitem [{\citenamefont {Goodman}(1981)}]{PIHalfSpace}%
  \BibitemOpen
  \bibfield  {author} {\bibinfo {author} {\bibfnamefont {M.}~\bibnamefont
  {Goodman}},\ }\bibfield  {title} {\bibinfo {title} {Path integral solution to
  the infinite square well},\ }\href {https://doi.org/10.1119/1.12720}
  {\bibfield  {journal} {\bibinfo  {journal} {American Journal of Physics}\
  }\textbf {\bibinfo {volume} {49}},\ \bibinfo {pages} {843} (\bibinfo {year}
  {1981})},\ \Eprint {https://arxiv.org/abs/https://doi.org/10.1119/1.12720}
  {https://doi.org/10.1119/1.12720} \BibitemShut {NoStop}%
\end{thebibliography}%

\end{document}